\documentclass[12pt,preprint,numberedappendix]{aastex}
\usepackage{graphicx, array, epsfig}% Include figure files
\usepackage{dcolumn}% Align table columns on decimal point
\usepackage{bm}% bold math
\usepackage{amssymb, subfloat, tabularx, rotating} %adjustbox, rotating, tikz
\usepackage[caption=false]{subfig}
\usepackage{amsmath}
\usepackage{epsfig}    
\usepackage{color,multirow}
\usepackage{slashed}
\usepackage{hhline}
\def\beq{\begin{equation}}
\def\eeq{\end{equation}}
\newcommand{\bea}{\begin{eqnarray}}
\newcommand{\eea}{\end{eqnarray}}
\newcommand{\Z}{\mathbb{Z}}

\numberwithin{equation}{section}
\usepackage{hyperref}

\makeatletter
\newcommand*{\myfnsymbolsingle}[1]{%
  \ensuremath{%
    \ifcase#1% 0
    \or % 1
      *%   
    \or % 2
      \dagger
    \or % 3  
      \ddagger
    \or % 4   
      \mathsection
    \or % 5
      \mathparagraph
    \else % >= 6
      \@ctrerr  
    \fi
  }%   
}   
\makeatother

\usepackage{alphalph}
\newalphalph{\myfnsymbolmult}[mult]{\myfnsymbolsingle}{}

\begin{document}
%%%%%%%%%
\title{\Large {Confronting Galactic and Extragalactic $\gamma$-ray observed by Fermi-LAT with Annihilating Dark Matter in Inert Higgs Doublet Model}}
\vspace {0.5cm}
\author{\bf Kamakshya Prasad Modak~\footnote{\bf kamakshya.modak@saha.ac.in}, Debasish Majumdar~\footnote{\bf debasish.majumdar@saha.ac.in}}
\affil{Astroparticle Physics and Cosmology Division, \\ Saha Institute of Nuclear Physics, \\ Kolkata 700064, India}
%\email{kamakshya.modak@saha.ac.in}
%\email{debasish.majumdar@saha.ac.in}

%\vspace{2.0cm}

\begin{abstract}

In this thorough study we focus on the indirect detection of Dark Matter (DM) through 
the confrontation of unexplained galactic and extragalactic $\gamma$-ray signatures for
a low mass DM model. For this, we consider a simple Higgs-portal DM model, namely, 
the inert Higgs doublet model (IHDM) where the Standard Model is extended with 
an additional complex SU(2)$_L$ doublet scalar. 
The stability of the DM candidate in this model, i.e., the lightest
neutral scalar component of the extra doublet, is ensured by imposing discrete $\Z_2$ symmetry.
The reduced-$\chi^2$ analysis with the theoretical, experimental and observational 
constraints suggests the best-fit value of DM mass in this model to be $\sim$ 63.5 GeV. 
We analyse the anomalous
GeV $\gamma$-ray excess from Galactic Centre 
in light of the best-fit IHDM parameters. We further check the consistency of the 
best-fit IHDM parameters with the Fermi-LAT obtained limits on photon flux for 
18 Milky Way dwarf spheroidal satellite galaxies (dSphs) known to be mostly dominated by DM.
Also since the $\gamma$-ray signal from DM annihilation is assumed
to be embedded within the extragalactic $\gamma$-ray background (EGB),
the theoretical calculations of photon flux for 
the best-fit parameter point in the IHDM framework 
are compared with the Fermi-LAT results for 
diffuse and isotropic EGB for different 
extragalactic and astrophysical background parametrisations. 
We show that the low mass DM in IHDM framework can satisfactorily confront 
all the observed continuum $\gamma$-ray fluxes originated from galactic 
as well as extragalactic sources.
The extensive analysis performed in this work is valid for any Higgs-portal model with DM mass 
in the ballpark of that considered in this work.

\end{abstract}

%\keywords{dark matter}
\maketitle

\newpage

\section{Introduction}
%%%%%%%%%%%%%%%%%%%%%%%%%%%%%%%%%%%%%%

The presence of dark matter (DM) in the universe is now established following
various astrophysical and cosmological~\citep{Ade:2013zuv, Begeman:1991iy, 
Komatsu:2010fb, Massey:2007wb} evidences. 
The recent data of PLANCK~\citep{Ade:2013zuv} suggest that $\sim$ 26.8\%
of the total mass-energy content of the universe consists of
cold (or non-relativistic) dark matter
whose particle nature is yet to be resolved. 
The weakly interacting massive particle, commonly known
as WIMP appears to be the most promising particle candidate 
of cold DM in the universe.

There are various ongoing experiments for the detection of 
dark matter both through direct and indirect mechanisms. 
In case of direct detection, the dark matter may scatter off 
a nucleus of a detecting material and in such direct detection 
experiments attempts are made to measure this recoil energy of the 
nucleus as the signature of dark matter detection. 
There are various ongoing direct detection experiments around the world 
such as  CDMS~II~\citep{Ahmed:2009zw, Agnese:2013cvt, Agnese:2013rvf}, 
CRESST~II~\citep{Angloher:2011uu}, CoGeNT~\citep{Aalseth:2010vx}, XENON~100~\citep{Aprile:2011hi, Aprile:2012nq}, 
LUX~\citep{Akerib:2013tjd} etc. that use different detection 
material, e.g. Ge, Si, Xe etc.  

The dark matter in the universe, because of its all pervading nature, 
may be trapped by 
very massive astrophysical objects such as
galactic centre,
solar core etc. and may undergo multiple scattering with the dense 
matter present at those sites losing in the process their velocity 
of escape and eventually are trapped inside these bodies. 
When accumulated in large numbers, these trapped dark matter 
particles may undergo pair annihilation to produce the pairs 
of standard model particles such as $q{\bar q}$ or $\ell {\bar {\ell}}$ 
as primary or secondary products. Gammas can be obtained as 
secondary products from the pair annihilation of these primary pairs 
of SM fermions (such as hadronisation of $b{\bar b}$ through $\pi^0$). 
The indirect dark matter search experiments look 
for these gamma rays or the other SM particles from dark matter 
pair annihilation.

The satellite borne detectors such as Fermi-LAT or 
Fermi Large Area Telescope detect the gamma rays from the the galactic 
centre (GC) and inner galaxy regions.
Any anomalous GeV gamma-ray excess from the Fermi-LAT observation
may indicate 
DM pair-annihilation at the galactic centre region in case
other known astrophysical phenomena fails to explain such excesses.
%%%%%%%%%%%%%%%%%%%%%%%%%%%%%%%%%%%%%%%%%%%%%%%%%%%%%%%%%%%%%%%%%%%%%
This excess gamma-ray signal %at the lower galactic latitude 
can be explained with DM models where DM particle annihilates
mainly into $q\bar{q}$ channels with the desired value of thermally 
averaged annihilation cross-section similar to the
canonical cross-section of typical thermal production of DM. 
%%%%%%%%%%%%%%%%%%%%%%%%%%%%%%%%%%%%%%%%%%%%%%%%%%%%%%%%%%%%%%%%%%%%%
This bump-like feature indicating gamma ray excess is 
also reported by Fermi-LAT collaboration~\citep{simonatalk} for the gamma rays from 
the galactic centre region. An early analysis of Fermi-LAT data reveals that
the gamma rays from the galactic centre region exhibit excesses (bump)
in the energy range $\sim 0.3 - 10$ GeV~\citep{Hooper:2011ti, Hooper:2012ft, Hooper:2013rwa, Huang:2013pda}. 
More involved and modified recent analysis~\citep{Daylan:2014rsa} including more recent 
data restricts the range of excess gamma to be in the energy region of 
$\sim 1 - 3$ GeV. 

The early analyses~\citep{Hooper:2011ti} by Dan Hooper {\it et al.} suggest that the
gamma-ray excesses mentioned above and the morphological 
features of these excesses can be satisfactorily explained by 
considering DM particles in the mass range of 
$\sim 30-60$~GeV annihilating only through $b{\bar b}$ channel. 
From the latter analysis~\citep{Hooper:2013rwa, Huang:2013pda}, the authors 
also gave the best fit value    
of the dark matter mass to be $61.8^{+6.9}_{-4.9}$~GeV annihilating only into 
$b \bar{b}$ pair with thermally 
averaged annihilation cross-section 
$\langle\sigma v\rangle = 3.30^{+0.69}_{-0.49}\times 10^{-26}$ cm$^3$s$^{-1}$ 
for explaining the above mentioned gamma ray excess. 
Another analysis~\citep{Hooper:2012ft} considering the dark matter to have primarily annihilated 
only to $\tau {\bar \tau}$, yields the dark matter   
mass in the range  $\sim 7-10$~GeV  
with thermally averaged annihilation cross-section 
$\langle\sigma v\rangle = 5.6 \times 10^{-27}$ cm$^3$s$^{-1}$. 
A very recent analyses~\citep{Daylan:2014rsa, Lacroix:2014eea} however 
indicate that a DM particle with mass
$\sim 31-40$~GeV and annihilating entirely into $b \bar{b}$ 
channel with thermally averaged annihilation cross-section 
$\langle\sigma v\rangle = (1.4-2.0) \times 10^{-26}$ cm$^3$s$^{-1}$ 
(normalised to a local DM density of 0.3 GeV/cm$^3$) 
can provide much better agreement with the nature of the low energy 
gamma-ray spectrum (with an excess in the energy range $\sim 1 - 3$
GeV). This new analyses also disfavour the
possibility of the previous proposition of $\sim$ 10 GeV DM annihilating solely 
into $\tau\bar{\tau}$ channels.~\footnote{cosmic ray positron data
also disfavour the possibility of DM annihilating into 
$\tau$ lepton pairs with the proposed mass of DM ($\sim 10$~GeV) 
and annihilation cross-section to accommodate GeV gamma-ray excess.} 
There are several attempts~\citep{
Logan:2010nw,
Buckley:2010ve,
Zhu:2011dz,
Marshall:2011mm,
Boucenna:2011hy,
Buckley:2011mm,
Hooper:2012cw,
Buckley:2013sca,
Anchordoqui:2013pta, Modak:2013jya, Guo:2014gra, Yu:2014pra, Cahill-Rowley:2014ora, 
Borah:2014ska, Banik:2014eda, Okada:2014usa, Cheung:2014lqa, Basak:2014sza, 
Berlin:2014pya, Ghosh:2014pwa, Ko:2014gha, Balazs:2014jla, Agrawal:2014una,
Agrawal:2014aoa, Izaguirre:2014vva, Cerdeno:2014cda, Ipek:2014gua, Boehm:2014bia, 
Wang:2014elb, Fields:2014pia, Arina:2014yna, Huang:2014cla,
Ko:2014loa, Agrawal:2014oha, Okada:2013bna, Boehm:2014hva,
Alves:2014yha, Berlin:2014tja, Abdullah:2014lla, Martin:2014sxa, Cline:2014dwa, Detmold:2014qqa,
Chang:2014lxa, Bell:2014xta, Cao:2014efa, Freytsis:2014sua, Heikinheimo:2014xza, Agashe:2014yua, Ghorbani:2014gka, Cerdeno:2015ega} 
to propose DM models and studying 
various aspects. In a more recent analyses~\citep{Calore:2014nla, Calore:2014xka, simonatalk, Agrawal:2014oha}
the galactic centre excess has been reanalysed considering several distinct
galactic diffuse emission models and the allowed DM mass range for the generation 
of such galactic centre gamma ray excess is severely relaxed. 
The preferred mass range of DM annihilating solely to $b\bar{b}$ 
channel is derived to be $35-165$ GeV~\citep{Agrawal:2014oha}.
In Ref.~\citep{Calore:2014xka} the Fermi-LAT galactic centre
GeV excess is interpreted with DM mass allowed up to 74 GeV 
with $b\bar{b}$ annihilation channel.
Alternative propositions other than annihilating DM
explanations such as unresolved millisecond pulsars to be the 
main origin of this observed gamma-ray excess have been
discarded since the observed anomalous gamma-ray emission 
extends much beyond the central stellar cluster.

Besides galactic centre, the dwarf spheroidal galaxies or dSphs
also are very rich in dark matter. 
These are faint companion galaxies of Milky Way. 
The mass to luminosity ratio ($\frac {M} {L}$) are found to be much higher than
$\left | \frac {M} {L} \right |_\odot$ where 
$\left | \frac {M} {L} \right |_\odot$ denotes the mass to luminosity 
ratio of the sun indicating that these are rich in dark matter.
These dark matter can pain annihilate and emit $\gamma$-rays.
With the wealth of Fermi-LAT $\gamma$-ray data, much detailed and 
thorough analyses~\citep{Abdo:2010ex, Ackermann:2011wa, GeringerSameth:2011iw, Mazziotta:2012ux,
GeringerSameth:2012sr, Ackermann:2013yva} performed on several dSphs to constrain DM annihilation. 

Apart from the galactic cases, the observed $\gamma$-ray signal 
by Fermi-LAT from the extragalactic sources
also may contain the signature of the dark matter annihilation at extragalactic
sites~\citep{Ullio:2002pj, Bergstrom:2001jj,
Gao:1991rz, Stecker:1978du, Taylor:2002zd, Ng:2013xha}. 
The signal may also have embedded in it the $\gamma$-ray from other possible
effects other than DM annihilation. To this end, there are 
attempts~\citep{Calore:2013yia, Cholis:2013ena, Tavakoli:2013zva, Sefusatti:2014vha, Ajello:2015mfa,
DiMauro:2015tfa, DiMauro:2015ika, Ackermann:2015tah} to 
extract DM signal from such extragalactic $\gamma$-ray background (EGB) and to provide
limits on DM annihilation cross sections for different DM masses. 
This requires proper modelling of the extragalactic parameters as well as 
proper knowledge about the other astrophysical backgrounds that
contribute dominantly to the EGB signal. 
With the analyses of new data collected by Fermi-LAT mission, a much 
detailed and clear picture of EGB has been put forward. The information 
regarding the astrophysical sources such as BL Lacs, millisecond pulsars,
star forming galaxies, radio galaxies etc. which possibly contribute to this EGB
are unveiled from various observations in radio and gamma wavelengths.
As Fermi-LAT collects more data, one 
can precisely measure the EGB spectra and put stringent constraints on
DM annihilation cross section. This constraints are contemporary to that obtained
from dSphs and galactic centre regions and may, in principle, put 
limits on various DM models in future.

A number of particle physics models for the dark matter candidate 
has been proposed and studied in literature. They include  
various extensions of Standard Model~\citep{Cheng:2002ej, Servant:2002aq, Duffy:2009ig, Randall:1998uk, Ma:2006km} 
whose DM phenomenologies~\citep{Modak:2012wk, Modak:2014vva} are
explored at length. Amongst them, the Higgs-portal models such as
singlet scalar DM~\citep{Silveira:1985rk, Veltman:1989vw, Burgess:2000yq, Barger:2007im, Gonderinger:2009jp}, 
inert Higgs doublet model (IHDM)~\citep{Deshpande:1977rw, LopezHonorez:2010tb} and  
two Higgs doublet model (2HDM)~\citep{Branco:2011iw}, singlet vector DM~\citep{Kanemura:2010sh, Lebedev:2011iq, Djouadi:2011aa}, 
singlet fermionic DM~\citep{Kim:2008pp} could be 
of particular interest for the present scenario in explaining the 
observed anomalous gamma emission by Fermi-LAT. 
The Higgs-portal models are interesting to study since the 
low mass DM candidates
of these models annihilate into quark pairs with the cross-section 
in the right ball park of thermal production. 
A special feature of these types of
models is that the DM candidates in these models 
exhibit resonance phenomena when their masses reach the value of $\sim$ half of the Higgs mass
while satisfying bounds given by PLANCK experiment (relic density) 
and dark matter direct detection experiments.  

In this work, we focus on the Inert
Higgs Doublet model (IHDM), proposed by Deshpande and Ma~\citep{Deshpande:1977rw} and confront 
the recently observed gamma-ray excesses from 
galactic centre region %and Fermi Bubbles 
with the dark matter candidate
in this model. We also explore the possibilities of 
the observation of gamma rays from 25 dwarf spheroidal galaxies
by this IHDM dark matter candidate.  
%%%%%%%%%%
In addition, we study the extragalactic gamma ray signals 
obtained by Fermi-LAT with this inert Higgs doublet dark matter. 
In the inert Higgs doublet model or IHDM an extra scalar doublet is 
added to the Standard Model (SM)
which is assumed to develope zero vacuum expectation value after 
spontaneous symmetry breaking. 
The model has been extensively studied in the context of both 
collider and DM phenomenology~\citep{Goudelis:2013uca, Gustafsson:2012aj,
LopezHonorez:2010tb, Hambye:2007vf, Agrawal:2008xz, Hambye:2009pw, Nezri:2009jd, 
Andreas:2009hj, Arina:2009um, Honorez:2010re, Melfo:2011ie, Krawczyk:2013jta, Gustafsson:2012aj, 
Dolle:2009fn, Kanemura:2011sj, Arhrib:2012ia, Swiezewska:2012eh, Swiezewska:2013uya, 
Lundstrom:2008ai, Cao:2007rm, Dolle:2009ft}. 
For the dark matter candidate in this IHDM framework, %(IHDM DM or lightest inert particle),   
a reduced $\chi^2$ analysis is performed considering all the above data and constraints 
and the best fit values 
for dark matter mass, annihilation cross-section and other parameters
of the model (various coupling constants) are found out.
In the present work we adopt those best fit values  
as the benchmark point and
study the Fermi-LAT gamma-ray flux results from 
both galactic (galactic centre, %Fermi Bubbles, 
dSphs) and extragalactic sources.

The paper is organised as follows. In Section~\ref{ihdm_sec} the theoretical 
framework of the IHDM 
is briefly described. Also the theoretical, observational and 
experimental constraints imposed on this model
are  discussed in this section. Confronting the observed gamma-ray 
excess from galactic centre in this model framework with a detailed study of 
the computed gamma-ray spectra is performed in Section~\ref{gc_sec}.
We compare the calculated results with the bin-by-bin upper 
limits on photon energy spectra for various Milky Way dSphs in Section~\ref{dsphs_sec}.
The Section~\ref{EGB_sec} contains the confrontation of the extragalactic $\gamma$-ray background
with calculated photon spectra in IHDM considering different extragalactic parametrisations
and astrophysical non-DM backgrounds.
In Section~\ref{sum_sec} we summarise our study and some important conclusions have been drawn.

%%#######################################################%%
\section{Inert Higgs Doublet Model (IHDM) framework\label{ihdm_sec}}
%%#######################################################%%

The Inert Higgs Doublet Model is 
one of the simplest extensions of Standard Model (SM)
Higgs sector where an additional complex scalar doublet, 
$\Phi$, odd under the discrete symmetry $\mathbb{Z}_2$, is considered 
alongwith the SM Higgs doublet, $H_1$.
After spontaneous symmetry breaking, while the SM Higgs gets a vacuum
expectation value (vev) $v$, the additional doublet does not acquire any vev.
Thus under $\mathbb{Z}_2$ symmetry $\Phi \rightarrow - \Phi$ and $H \rightarrow H$
(even under $\mathbb{Z}_2$) and after symmetry breaking
the two doublets $H$ and $\Phi$ can be expanded as,
\begin{equation}
	H ~=~ \left( \begin{array}{c} G^+ \\ \frac{1}{\sqrt{2}}\left(v+h^0+\mathrm{i}G^0\right) \end{array} \right),
	\qquad
	\Phi ~=~ \left( \begin{array}{c} H^+\\ \frac{1}{\sqrt{2}}\left(H^0+\mathrm{i}A^0\right) \end{array} \right),
\end{equation}
where $G^{\pm}$ and $G^0$ are charged and neutral Goldstone bosons 
respectively. Note that vev of these scalar doublet fields are 
$\langle H\rangle = v/\sqrt{2}$ ($v \simeq 246$~GeV) and $\langle \Phi\rangle = 0$.

With the unbroken
$\mathbb{Z}_2$ symmetry the model has a CP-even neutral scalar $H^0$, 
a CP-odd neutral scalar $A^0$, and a pair of charged
scalars $H^{\pm}$. Since the $\mathbb{Z}_2$ symmetry 
excludes the couplings of fermions with $H^0$, $A^0$, $H^{\pm}$,
the decay of the latter particles to fermions are thus prevented.
This ensures the stability of lightest neutral 
scalar ($H^0$ or $A^0$) and hence the lightest among these two 
can serve as a possible DM candidate. Either $H^0$ or $A^0$ is chosen as 
the lightest inert particle or LIP and is the candidate of dark matter
in the present model. 

The most general tree-level scalar potential of IHDM 
consistent with imposed $\mathbb{Z}_2$ symmetry can
be written as,
\begin{equation}
	V_0 ~=~ \mu_1^2 |H|^2  + \mu_2^2|\Phi|^2 + \lambda_1 |H|^4+ \lambda_2 |\Phi|^4 + \lambda_3 |H|^2| \Phi|^2
		+ \lambda_4 |H^\dagger\Phi|^2 + \frac{\lambda_5}{2} \Bigl[ (H^\dagger\Phi)^2 + \mathrm{h.c.} \Bigr]\,\,\, ,
\label{Eq:TreePotential}
\end{equation}
where $\mu_i$s and $\lambda_i$s denote various coupling parameters.
The model has set of six parameters, namely
\begin{equation}
 	\left\{ M_{h^0}, ~~ M_{H^0}, ~~ M_{A^0}, ~~ M_{H^{\pm}}, ~~ \lambda_L, ~~ \lambda_2 \right\},
	\label{eq:masses}
\end{equation} 
where $M_{h^0}, ~~ M_{H^0}, ~~ M_{A^0}, ~~ M_{H^{\pm}}$ are the masses of Higgs $h$, CP-even scalar $H^0$,
pseudo-scalar $A^0$ and charged scalars $H^{\pm}$. $\lambda_L$ and $\lambda_S$ are in couplings given by,
\begin{eqnarray}
\lambda_L &=& \frac{1}{2} \left( \lambda_3 + \lambda_4 + \lambda_5 \right), \\
\lambda_S &=& \frac{1}{2} \left( \lambda_3 + \lambda_4 - \lambda_5 \right).
\end{eqnarray}
The parameters $\lambda_{L}$ or $\lambda_{S}$ denote 
the coupling strength for $H^0H^0h^0$ (if $H^0$ is considered to be 
the lightest inert particle or LIP) or $A^0A^0h^0$ (if $A^0$ is the LIP). 
%The self-quartic coupling $\lambda_2$ does not have any viable contribution 
%as far as the tree-level processes are concerned.

%%%%%%%%%%%*******************************%%%%%%%%%%%%%%%%%%%%%

A detailed study of this model had been done in Ref.~\citep{Arhrib:2013ela} 
where the authors made use of the various constraints available 
from DM experiments and other results and made a $\chi^2$ 
analysis with the IHDM theory of the dark matter mentioned above.
Considering all possible experimental and theoretical constraints such as
Planck limits, direct detection constraints, unitarity, perturbativity, etc. as also
constraints from LHC, they provide the best fit values of the quantities 
$M_{h^0}$, $M_{H^\pm}$, $M_{H^0}$, $M_{A^0}$ and the parameters $\lambda_L$, 
$\lambda_2$ (shown in Table~\ref{table1}). 

\section{Confronting Gamma Ray flux from Galactic Centre %and Fermi Bubble 
in this framework \label{gc_sec}}
%%#######################################################%%

The differential gamma-ray flux due
to the annihilating DM coming from the galactic DM halo per unit 
solid angle can be written as~\citep{Cirelli:2010xx},
\begin{eqnarray}
\frac{d\Phi_{\gamma}}{d\Omega dE_{\gamma}} = \frac{1}{8\pi\alpha}
\sum_f\frac{\langle \sigma v\rangle_f}{M^2_{H^0, A^0}}
\frac{dN^{f}_{\gamma}}{dE_{\gamma}} r_{\odot} \rho^2_{\odot} J \,\, ,
\label{gammaflux1}
\end{eqnarray}
where $m_{H^0}$ is the mass of DM candidate $H^0$ 
and $\alpha=1$ for the present DM candidate (self-conjugated).
In Eq.~\ref{gammaflux1},
$r_{\odot}$ and $\rho_{\odot}$ are the distance of the solar system 
from the galactic centre and the local DM halo
density respectively. The factor $J$ in Eq.~\ref{gammaflux1} 
gives the total dark matter content at the target and is given by,
%can be expressed as,
%
\begin{eqnarray}
 J = \int_{l.o.s} \frac{ds}{r_\odot}
 \left(\frac{\rho(r)}{\rho_\odot}\right)^2 \,\,
 \label{j_lb_rt}
 \end{eqnarray}
where $\rho_{\odot}$ and $\rho(r)$ are respectively the DM density at solar region and
the density at a radial distance $r$ from GC and $r$ is expressed 
in terms of line of sight, $s$ as,
% with $r$ being the radial distance of the site of DM annihilation 
% at galactic centre neighbourhood 
% from the galactic centre and can be expressed
% in terms of line of sight, $s$ as,
\begin{eqnarray}
r = \begin{cases}
     \left( s^2 + r^2_{\odot} - 2sr_{\odot} {\rm cos}\,{l} \,
{\rm cos}\,{b}\right)^{1/2}\,\, &  (\text{galactic}\,\,\, l, b\,\,\,
\text{coordinate})\\
     \left( s^2 + r^2_{\odot} - 2sr_{\odot} {\cos}\,{\theta}\right)^{1/2}\,\, 
& (\text{galactic}\,\,\, r, \theta\,\,\,\text{coordinate})
    \end{cases}
\label{los}
\end{eqnarray}
In the above $\rho(r)$ is the DM halo profile and for a generalised 
NFW DM halo the analytical expression can be given~\citep{Navarro:1995iw, Navarro:1996gj} as,
\begin{equation}
  \rho(r) = \frac{\rho_{\odot}}{(r/r_c)^\gamma[1 + (r/r_c)^\gamma]^{(3.0 - \gamma)}} 
\,\,\, ,
 \label{gnfw}
 \end{equation}
%
% with $\rho_{\odot}$ being the local DM halo density at the solar location ($\rho(r_{\odot})$, $r_{\odot}$ ($\sim$ 8.5 kpc) is 
% the distance of sun from GC) chosen to be $\sim$ 0.4 GeV/cm$^3$. 
and $\rho_{\odot}$ $\sim$ 0.4 GeV/cm$^3$.
In the above relation, the values of the parameters
$r_c$ and $\gamma$ are taken to be 20 kpc and 1.26 respectively 
in the present calculation following Ref.~\citep{Daylan:2014rsa}. We use 
\texttt{micrOMEGAs}~\citep{Belanger:2010gh, Belanger:2013oya} code 
to compute various DM observables in this model.

\begin{figure}[cbt!]
\begin{center}
\includegraphics[angle=-90, scale=0.4]{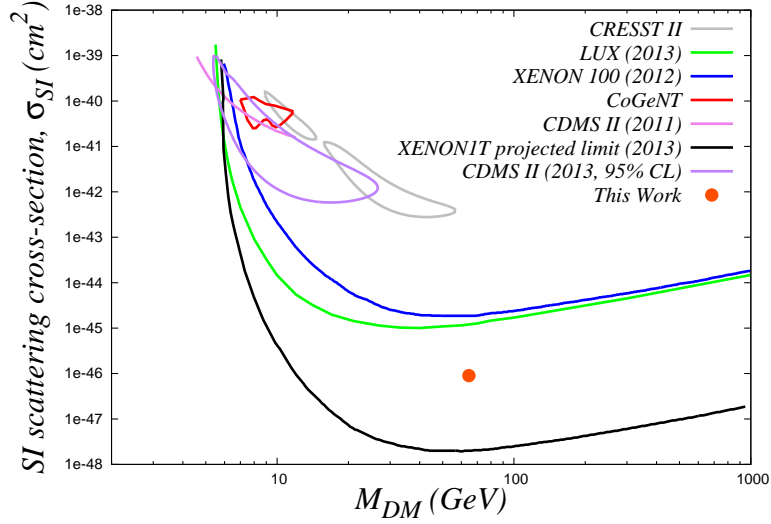}
%%%
   \caption{{\it Various DM direct detection experimental bounds on 
the benchmark point of Table~\ref{table1}. See text for details.}}
   \label{direct}
\end{center}
\end{figure}

The calculated branching ratio 
for the channel ${\rm LIP}~{\rm LIP} \to b\bar{b}$ is found to be 69.2\%.
The branching ratios for other annihilation channels 
in case of the present LIP dark matter are also computed 
and they are tabulated in Table~\ref{table2}.  
Since the DM mass is 
close to half of the SM Higgs ($\sim m_{h}/2$) like particle
discovered by LHC, we will have resonance
effect in obtaining the required cross-section
($2.37\times10^{-26}$ cm$^3$s$^{-1}$) with $b\bar{b}$ 
as dominant channel for the typical thermal production of DM. Also 
DM-nucleon scattering cross-section for the chosen benchmark 
point in this model is about $8.89\times10^{-11}$ pb 
(averaged value per nucleon for interaction with Xe nucleus)
which is just under the the present bounds for XENON~100 and LUX 
experiments. This is shown in Fig.~\ref{direct}. The future DM direct
searches like XENON1T~\citep{Aprile:2012zx} can easily probe this point as seen 
from Fig.~\ref{direct}.

The Fermi-LAT data for gamma-ray flux from the inner 
$5\degr$ surrounding of the galactic centre have been
studied in Ref.~\citep{Hooper:2012ft}. 
The known $\gamma$-ray sources
in this region can be found in (extracted from) 
Fermi Second Source Catalog (2FGL)~\citep{Fermi-LAT:2011iqa}.
Although Fermi Third Source Catalog (3FGL) has recently been made 
available \citep{TheFermi-LAT:2015hja},
but no analysis of the background or analysis for $\gamma$-ray from other known
processes at the region of interest has been reported yet.   
Also cosmic ray interaction
with gas distributed in this galactic region produces neutral 
pions that subsequently decay to produce
enormous amount of $\gamma$-ray. This is a viable mechanism for 
known disc template emission. 
Now, the spectral and morphological feature of the photon 
flux from inner $5\degr$ subtending the GC
after subtracting the contribution from both the known sources 
of Fermi Second Source Catalog and disc
template emission shows a `bumpy' nature
% , i.e., the photon count is higher for 
in $\gamma$-ray energy
ranging from $\sim$ 300 MeV to $\sim$ 10 GeV. The count 
drops significantly after 10 GeV of $\gamma$-ray
energy.

\begin{figure}[cbt!]
\begin{center}
\includegraphics[angle=-90, scale=0.4]{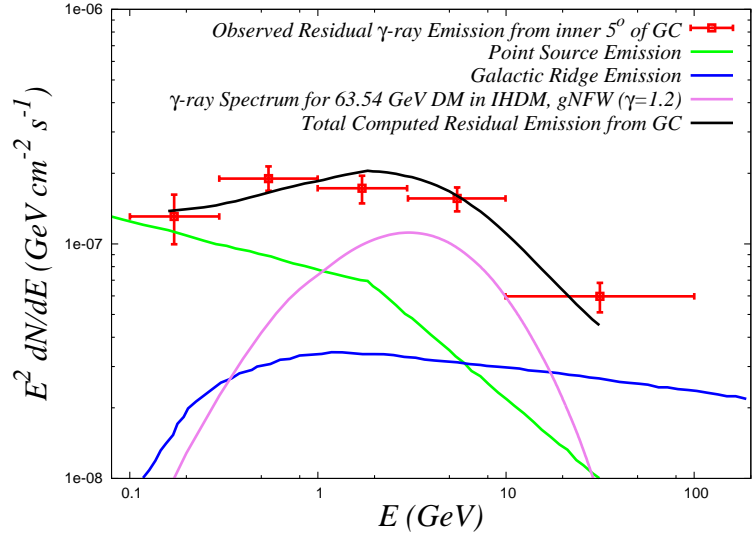}
%%%
   \caption{{\it Residual GeV gamma-ray flux from the inner $5\degr$ surrounding the galactic centre. 
   See text for details.}}
   \label{gc_5deg}
\end{center}
\end{figure}

We have computed the $\gamma$-ray spectrum from 
inner $5\degr$ subtending the GC
from DM annihilation within the present framework of inert Higgs doublet 
model for dark matter particle. The 
chosen benchmark points for the parameters of the model such as 
dark matter mass, coupling constants etc. are given in 
Table~\ref{table1}. The flux have been computed for the 
generalised NFW DM halo profile. 
Also included in the calculation are the contributions from both point 
source and galactic ridge emission. The total calculated
flux is then compared with the observed residual photon flux 
and the results are shown in Fig.~\ref{gc_5deg}.
In Fig.~\ref{gc_5deg}, the green-coloured and blue-coloured lines 
represent the fluxes for point source  
and galactic ridge emission respectively. 
The $\gamma$-ray spectrum
for DM annihilation for the benchmark points mentioned above, are shown
in purple line whereas the black line is for the total $\gamma$-ray flux
obtained by summing over all the fluxes represented by green, blue and purple
coloured lines in Fig.~\ref{gc_5deg}. For these calculations we have 
adopted the generalised NFW (gNFW) halo profile with $\gamma = 1.26$.  
The total flux (black line) is then compared with the observed 
residual emission data. These observed data are denoted by 
red-coloured points in this figure. It is clear from  Fig.~\ref{gc_5deg}
that our computation of total residual gamma emission (black line) 
agrees satisfactorily with the observational results. 

\begin{figure}[cbt!]
\begin{center}
\includegraphics[angle=-90, scale=0.32]{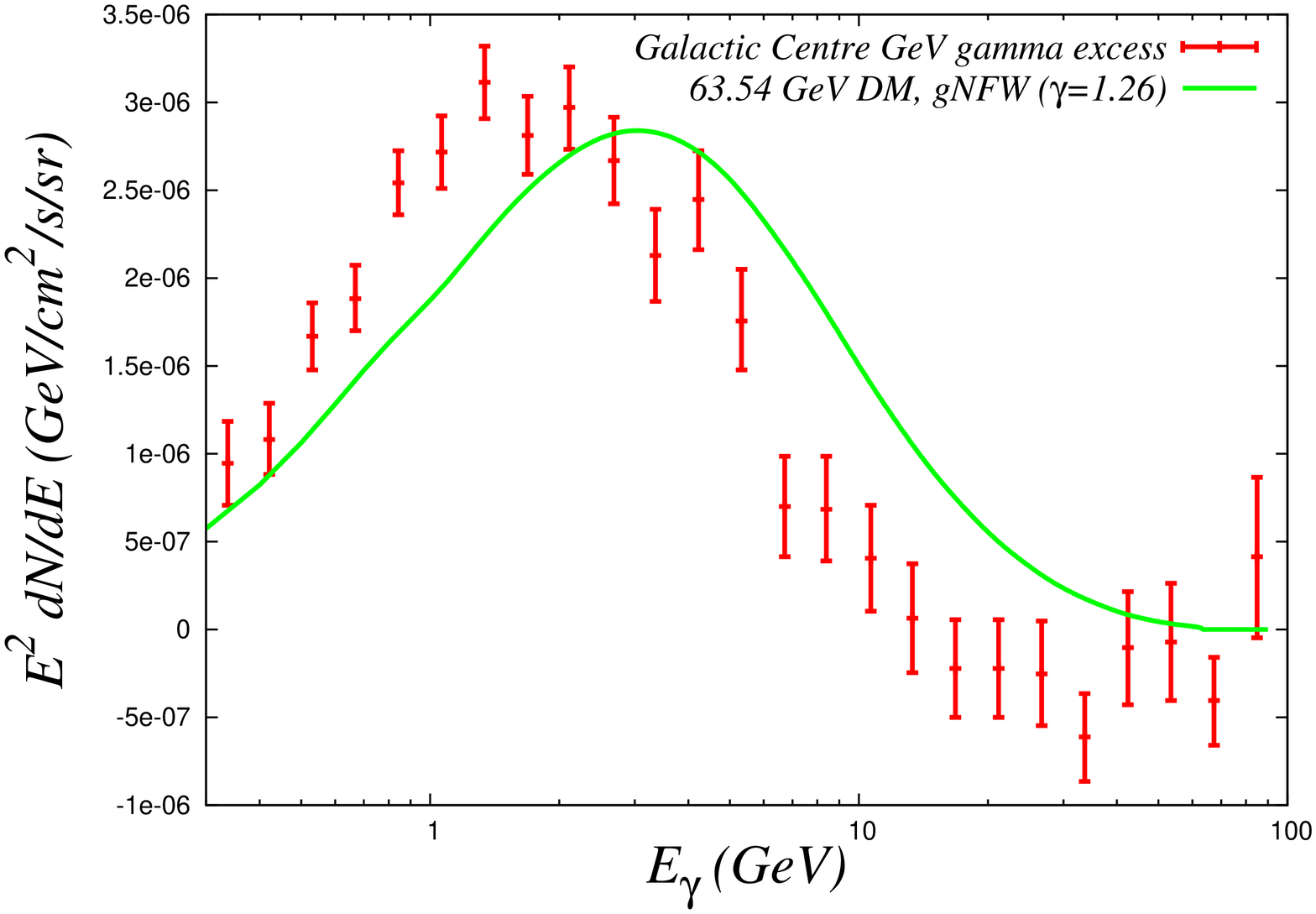}
%%%
\includegraphics[angle=-90, scale=0.32]{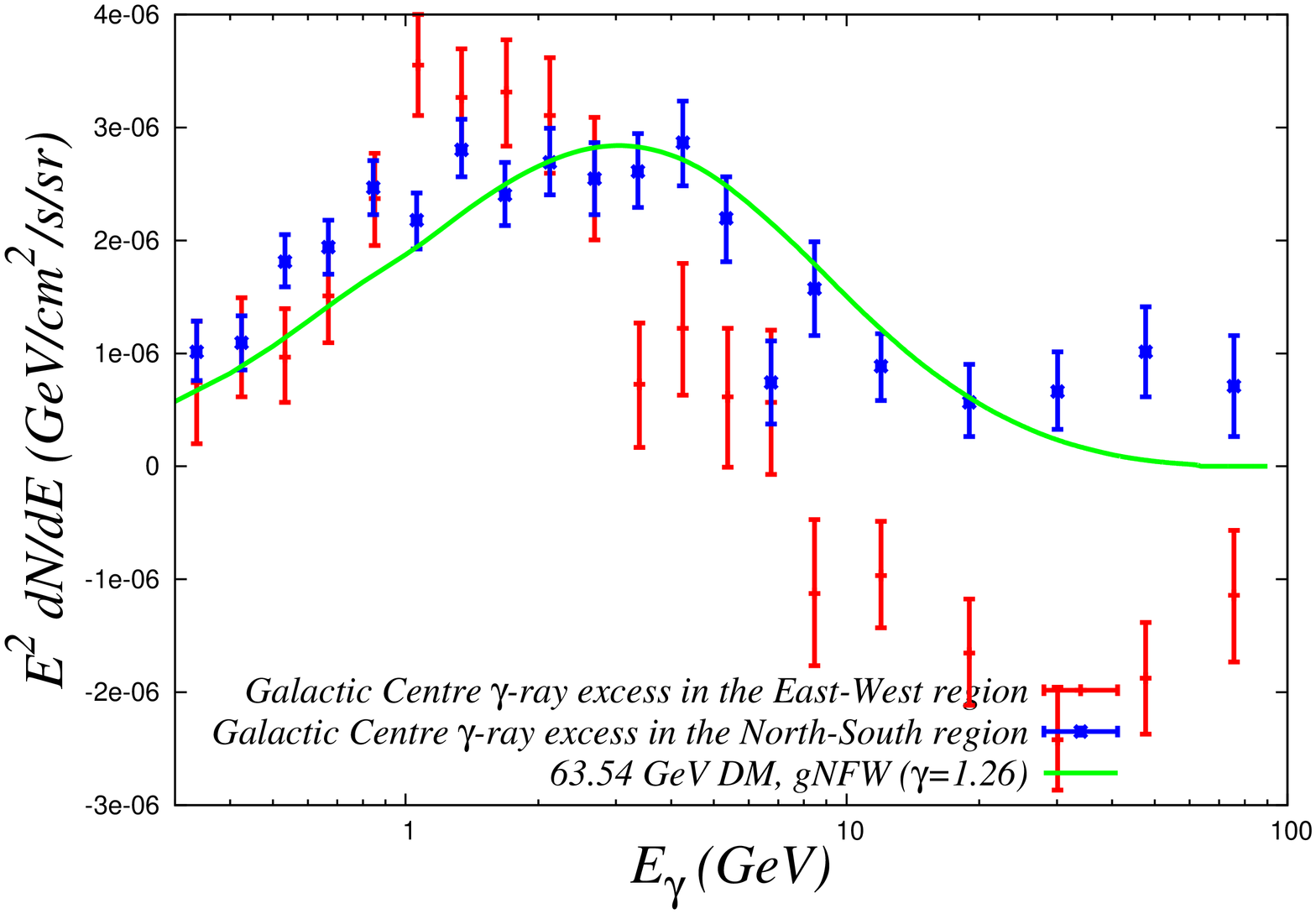}
\caption{{\it Left panel: Comparison of calculated $\gamma$-ray flux (green line; assuming DM
	annihilation) with the observed $\sim 1-3$ GeV gamma-ray excess from GC (red data points).
	Right panel: Same as the left panel but here the observed data are for 
	galactic `North-South' region ($|b|<|\ell|$; red data points) and for 
	galactic `East-West' region ($|b|>|\ell|$; blue data points). See text for details.
	}}
\label{gce}
\end{center}
\end{figure}

In a recent analysis of $\gamma$-ray flux from GC region 
where the analysis is only for the GC gamma ray (subtracting 
all possible contributions from other known astrophysical sources),
\citep{Daylan:2014rsa} an excess of gamma ray in the gamma energy region of  
$\sim 1-3$ GeV has been reported. This excess is shown in the left panel of Fig.~\ref{gce} with the 
red-coloured points. It is suggested in the same analysis that 
in order to explain this anomalous gamma ray excess from dark matter 
annihilation scenario, the DM mass should be in the range of 
31-40 GeV which is to annihilate purely into $b\bar{b}$ pair.
We calculated these gamma ray fluxes in our framework of inert Higgs doublet 
dark matter for the dark matter mass of $\sim$ 63.5 GeV (adopted from  
Table~\ref{table1} (benchmark point)) and compared our results with the 
analysed data points mentioned above
(red coloured points shown in the left panel of Fig.~\ref{gce}). 
In the left panel of Fig.~\ref{gce}, the green coloured 
line represents the present calculation. These calculations are 
performed considering gNFW halo profile with the halo parameter
$\gamma = 1.26$. It is to be noted from the left panel of Fig.~\ref{gce} 
that although the morphological 
feature of the spectrum from our calculation (green line) 
is similar in nature to that obtained 
from the analysis of Fermi-LAT observational data (red-coloured points)
from GC, 
the position of the  
maxima of excess gamma ray in our calculation is shifted to 
somewhat higher energy at $\sim 3.1$ GeV instead of being within 
the expected  energy range of $\sim 1-3$ GeV as obtained from the Fermi-LAT data.
However the calculated 
position of the peak (green line) is at $\sim 2.84 \times 10^{-6}$ 
GeV/cm$^2$/s/sr  which 
is in the same ballpark of the observed peak (red points).

More detailed analysis of the observed $\gamma$-ray excess
reveals~\citep{Daylan:2014rsa} a further anomaly between the gamma ray spectra 
for the gamma rays from galactic East-West region and from 
galactic North-South region. 
The North-South region is designated as $|b|<|\ell|$, where $b$ and $\ell$ are
the galactic latitude and longitude respectively and for the East-West 
region, $|b|>|\ell|$.
The two spectra are shown in the right panel of Fig.~\ref{gce}.
The red coloured points are for gamma spectrum from galactic East-West
region while the blue coloured points represent the spectrum from 
North-South region. As can be seen from the right panel of Fig.~\ref{gce}, the gamma flux 
from the present calculation (shown by green line in the right panel of Fig.~\ref{gce}) 
agrees more satisfactorily with the 
``North-South" gamma emission spectrum than the ``East-West" spectrum. 

%%%%%%%%%%%%%% CCW, FERMI, ABFH %%%%%%%%%%%%%%%%%%%%%%%%%%%%%%%%%%%%%%%%%%%%%%%%%%%%%%%%%%%
% Although the galactic centre $\gamma$-ray excess is statistically significant, it depends
% on the choice of diffuse background model that characterises the diffuse 
% emission over the entire sky. But for the galactic centre, it is very hard to
% get best suited model.
The systematic uncertainty for the estimation of background model provided by the Fermi-LAT 
is very large compared to the statistical uncertainty. 
Attempts~\citep{Calore:2014xka, simonatalk} are made to quantify such systematics
for galactic central region. 
We confront the flux obtained for our IHDM benchmark scenario (Table~\ref{table1}) 
using these two complementary approaches. 
% Since the dark matter distribution near the galactic centre is not known with certainty, the 
% proper estimation of $J$-factor in Eq.~\ref{j_lb_rt} is also difficult. 
We adopt the proposed method of Ref.~\citep{Agrawal:2014oha} for the uncertainty 
estimation of $J$-factor. 
% The uncertainty
% in the halo profile may arise from the two factors, namely from the index factor of dark
% matter distribution profile $\gamma$ and the local dark matter density $\rho_{\odot}$ 
% at a distance $r = r_{\odot} = 8.5$ kpc. 
This involves estimation of halo profile uncertainty and the uncertainty in local DM density.
The density profile index $\gamma$, is estimated to be 
$\gamma = 1.2\pm 0.1$ from different galactic
diffuse emission models and $\rho_{\odot}$ is estimated to be $\rho_{\odot} = 0.4\pm0.2$ 
GeV/cm$^{3}$ for different normalisation of halo.% these can introduce uncertainty in the astrophysical $J$-factor.
%To parametrised the uncertainty in the $J$-factor the following quantity is used, 
Defining the $J$-factor as
\begin{align}
\bar{J}
&=
\frac{1}{\Delta \Omega}
\int_{\Delta \Omega} J(\psi) d\Omega
\equiv
\mathbb{J}\times
\bar{J}_{\rm canonical} \,,
\end{align}
where $\bar{J}_{\rm canonical}$ is the central value of 
$\bar{J}$ and the factor $\mathbb{J}$ signifies the deviation from
the canonical halo profile due to the uncertainties of the profile. 
In the above $\Delta \Omega$ is the region of interest (ROI) 
for a given analysis. The astrophysical factor $J(\psi)$ is same as that
in Eq.~\ref{j_lb_rt}.

In Ref.~\citep{Calore:2014xka} a thoroughly analysis of Fermi-LAT data in the
inner galaxy region over the photon energy ranging from 300 MeV -- 500 GeV 
has been made where the chosen region of interest (ROI) is extended to a
$20^\circ\times20^\circ$ ($-20^\circ<\ell<20^\circ$, $-20^\circ<b<20^\circ$) square region
surrounding the galactic centre. The inner galactic latitude of $2^\circ$ 
($|b|<2^\circ$) has been masked out. 
For the canonical halo profile with
$r_\odot=8.5$ kpc, $\rho_\odot = 0.4$ GeV/cm$^3$, $r_c =
20$ kpc and $\gamma = 1.20$. The numerical value of $\bar{J}_{\rm canonical}$ is found
out to be $2.0\times 10^{23}$ GeV$^2$ cm$^{-5}$ for the canonical halo. 
%By studying a large number (60) of background models for galactic diffuse
%emission and the correlations in the $\gamma$-ray spectrum along the galactic disk containing
%very faint signal, the residual emission signal and the systematics uncertainties are extracted. 
The uncertainty ($\mathbb{J}\in[0.19,3]$) in the $\bar{J}$ factor is found out to be
in their analysis. 

\begin{figure}[cbt!]
\begin{center}
\includegraphics[angle=-90, scale=0.32]{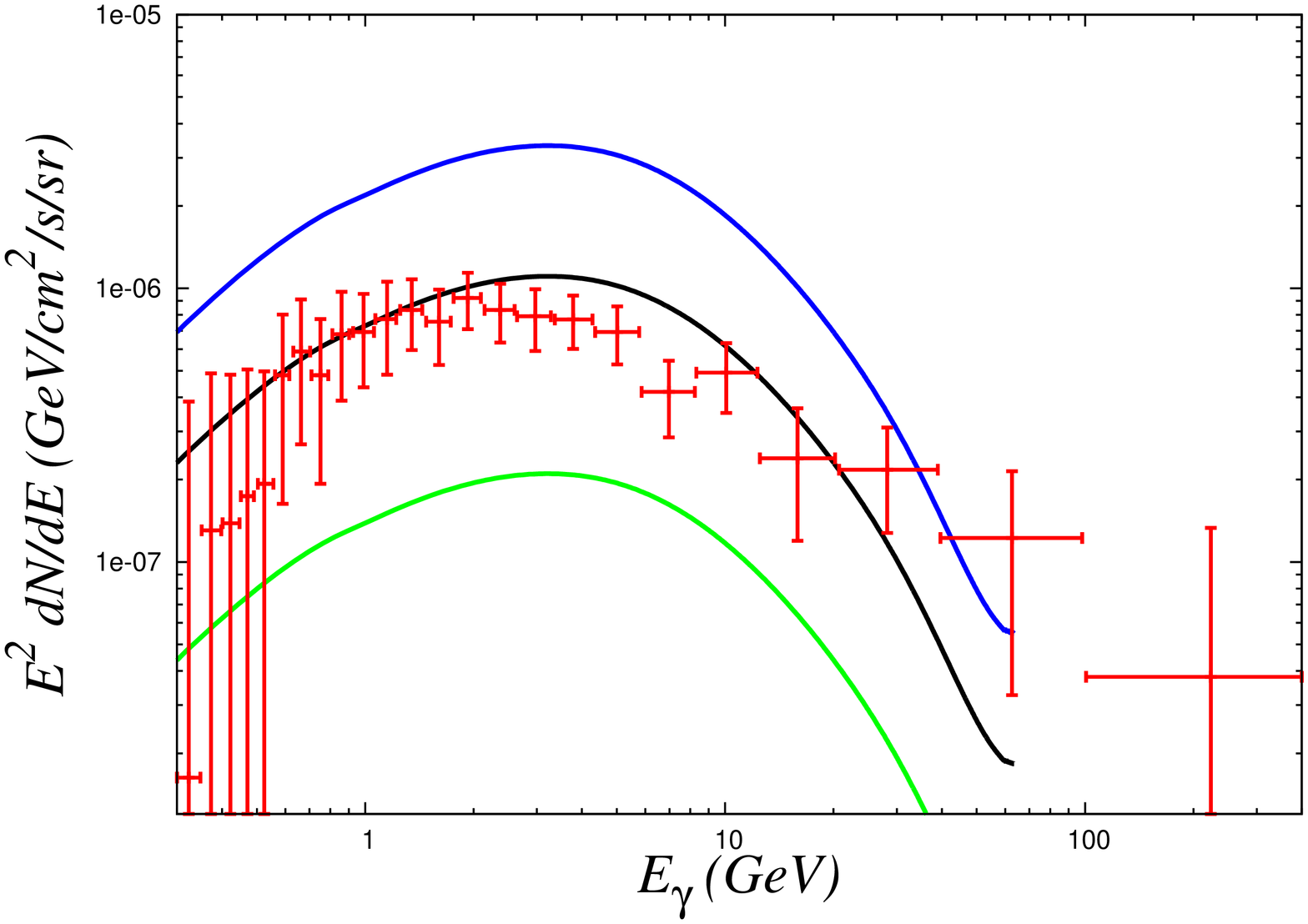}
\includegraphics[angle=-90, scale=0.32]{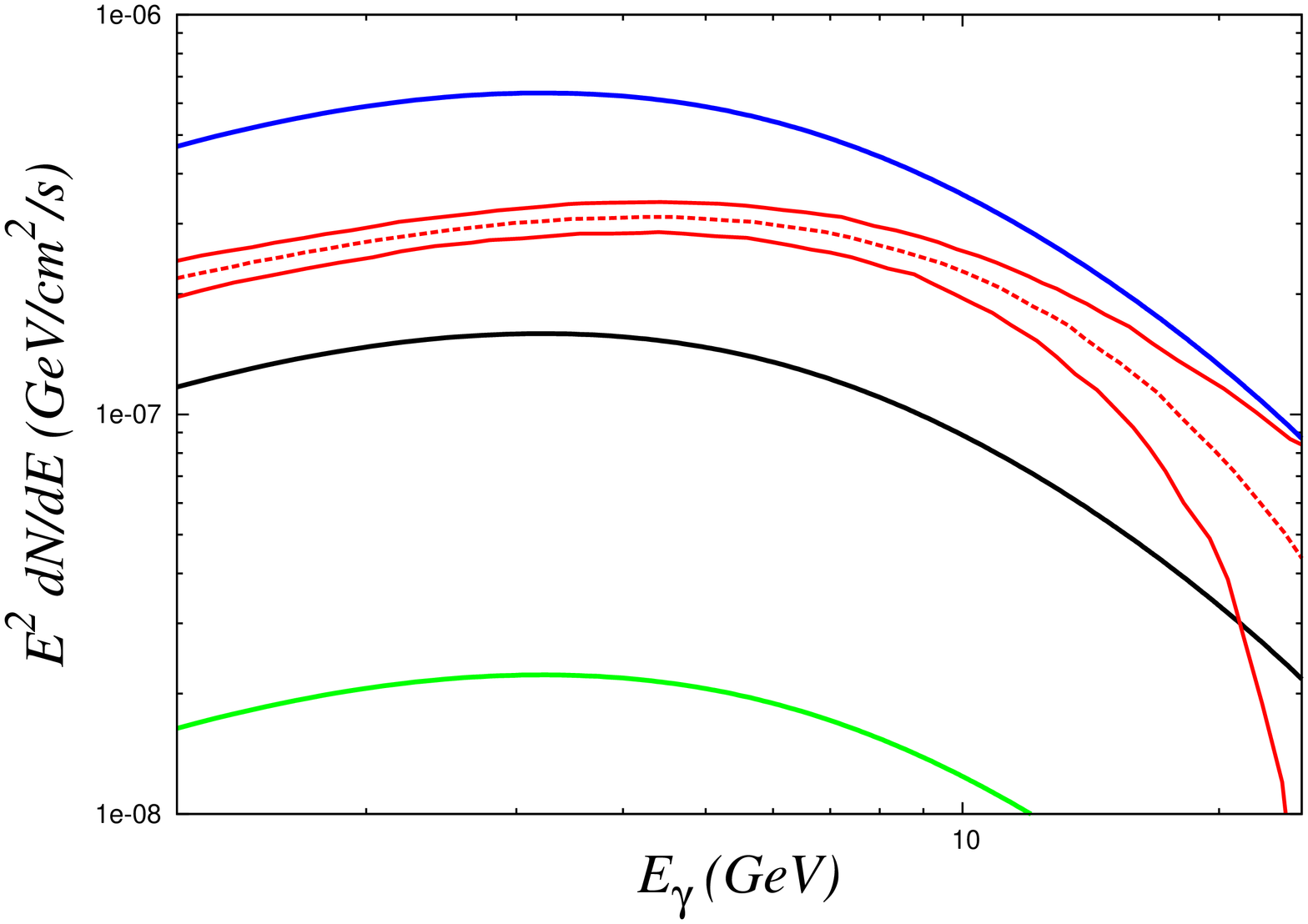}
%%%
\caption{{\it Left panel: Comparison of calculated $\gamma$-ray flux (red points) with
	the observed residual $\gamma$-ray spectrum for the region in galactic coordinate
	$|\ell|<20^\circ$, $2^\circ<|b|<20^\circ$. The black line is for canonical $J$-factor
	while green and blue lines are calculated with minimum and maximum deviations of $J$-factor
	from its canonical value. Right panel: Comparison of the calculated results in the left panel
	with observed $\gamma$-ray spectra (red points) obtained by studying $15^\circ\times15^\circ$
	region around GC. See text for details.}}
\label{ccw_agfh}
\end{center}
\end{figure}

% In order to have an estimation of the galactic $\gamma$-ray excess from the annihilation
% of dark matter in such parametrisation, we consider the LIP dark matter in IHDM
% framework and study the photon spectrum produced from its annihilation. 
Plugging in all of the canonical parameters for the dark matter halo model,
we compute the canonical $\gamma$-ray flux obtained from annihilation channels of $\sim$ 63.5 GeV
LIP DM (adopted from Table~\ref{table2} (benchmark point)). The resulting plot is shown in the left
panel of Fig.~\ref{ccw_agfh}
by black coloured line. The black line is obtained considering only the canonical value
$\mathbb{J}\in[0.19,3]$ of the $J$-factor. The red coloured points denote the residual spectrum of 
$\gamma$-ray excess in the galactic centre with highly correlated errors obtained from Ref.~\citep{Calore:2014xka}. 
We repeat the calculations by also taking the uncertainties
$\mathbb{J}\in[0.19,3]$ in the $\bar{J}$ factor. The blue and the green coloured lines 
in the left panel of Fig.~\ref{ccw_agfh} denote
the galactic $\gamma$-ray flux from $\sim$ 63.5 GeV DM annihilation for the 
maximum value of the uncertainty $\mathbb{J}_{\rm max} \sim 3.0$ and the minimum value of the 
uncertainty $\mathbb{J}_{\rm min} \sim 0.19$ respectively. 
% We can conclude 
From the left panel of Fig.~\ref{ccw_agfh} one sees
that the uncertainty factor $\mathbb{J}$ needs to be smaller than unity to have a better
fit to the data. %for 63.54 GeV LIP dark matter. 

On the other hand, recently the Fermi collaboration~\citep{simonatalk} has also studied the region 
surrounding the galactic centre using different background models of galactic diffuse
emission. The fit to the galactic centre $\gamma$-ray data is observed to improve
very significantly when an additional contribution similar to that from dark matter
annihilation is added. The Fermi collaboration has chosen a different 
region ($15^\circ\times15^\circ$) surrounding the galactic centre smaller 
than that chosen by Ref.~\citep{Calore:2014xka}. However for their analysis,
unlike in Ref.~\citep{Calore:2014xka}, in this case, the galactic centre is not masked out. 
Based on their preliminary
analysis the Fermi collaboration
has reported four best fit $\gamma$-ray spectra for the four distinct choices of background models
for galactic diffuse emission. The nature of these four best fit $\gamma$-ray spectra
differ notably after a few GeV photon energy.
The obtained $\gamma$-ray spectra is found to 
yield much more conservative measurement of the systematic uncertainty. Although
the Fermi has analysed the data using NFW halo profiles with slope values of 1.0 and 1.2,
Ref.~\citep{Agrawal:2014oha} has chosen a more conservative approach on $\gamma$ factor and set 
it to $1.2\pm0.1$. Also the value of $\rho_{\odot}$ is chosen to be $0.4\pm0.2$ GeV/cm$^{3}$.
Keeping the parameters of the $J$-factor fixed, one would obtain in this case, $\bar{J}_{\rm canonical}
= 1.58\times10^{24}$ GeV$^2$ cm$^{-5}$. The uncertainty in the $\bar{J}$ factor
as obtained from such parametrisation is $\mathbb{J}\in[0.14,4.0]$ for the 
Fermi analysis.

We make an estimation of the galactic $\gamma$-ray excess from the annihilation 
of dark matter as chosen in this work (benchmark point in Table~\ref{table1}) and 
confront the results with the $\gamma$-ray
spectra obtained by Fermi collaboration. We make the comparison of the calculated spectrum
with that reported by Fermi after preliminary analysis and show it in the right panel of Fig.~\ref{ccw_agfh}.
The black line in the right panel of Fig.~\ref{ccw_agfh} denotes the 
photon flux obtained using the canonical parametrisation of the halo
profile (with $\mathbb{J}=1$). The blue and the green coloured lines in the 
right panel of Fig.~\ref{ccw_agfh} represent
the galactic $\gamma$-ray flux %from the annihilation of 63.54 GeV DM 
calculated for the 
maximum uncertainty $\mathbb{J}_{\rm max} \sim 4.0$ and the minimum 
uncertainty $\mathbb{J}_{\rm min} \sim 0.14$ respectively. 
%The red coloured line denote the 
%upper and the lower limits of the photon flux from the galactic centre 
%as provided by Fermi in one of the four obtained spectra.
Also shown in the right panel of Fig.~\ref{ccw_agfh} the Fermi analysis results with its upper
and lower bounds.~\footnote{Out of the four spectra 
given by the preliminary analysis of Fermi, we choose only a particular spectrum
which provides the best fittings for dark matters with low masses since our interest in this paper
is on the low mass region in IHDM.} From the right panel of Fig.~\ref{ccw_agfh} it appears that 
unlike the previous analysis (with $-20^\circ<\ell, b<20^\circ$) the uncertainty factor $\mathbb{J}$ should be more than unity
in order to provide a better fit to the data for the considered IHDM benchmark point.

%%#######################################################%%
\section{Confronting Gamma-ray flux from Dwarf Spheroidal Galaxies 
in this framework \label{dsphs_sec}}
%%#######################################################%%

One of the most promising targets for the search of dark matter 
via indirect detection ($\gamma$-ray) is the dwarf spheroidal galaxies
(dSphs) of the Milky Way. The dSphs are considered to be promising
for the study of DM phenomenology because of their proximity, low astrophysical
background and huge amount of DM content.

The satellite-bourne gamma-ray experiment, Fermi-LAT, 
search for the
$\gamma$-ray sky in the energy range spanning from $\sim$ 500 MeV --
500 GeV~\citep{Atwood:2009ez}. 
In a more recent study by Ackermann {\it et al.}
~\citep{Ackermann:2013yva}, 4-year gamma ray data of Fermi-LAT 
on dSphs (04-08-2008 to 04-08-2012) with energy ranging from 500 MeV
to 500 GeV have been chosen for studying 25 independent Milky
Way dSphs galaxies.
The chosen dSphs galaxies are Bootes I, Bootes II, Bootes III,
Canes Venatici I, Canes Venatici II, Canis Major, Carina, Coma Berenices,
Draco, Fornax, Hercules, Leo I, Leo II, Leo IV, Leo V, Pisces II,
Sagittarius, Sculptor, Segue 1, Segue 2, Sextans, Ursa Major I,
Ursa Major II, Ursa Minor and Willman 1. The galactic coordinates 
as well as the 
radial distances from the galactic centre of
these dwarf galaxies are tabulated in Table~\ref{table3}.
From the analysis of their data, they set robust upper limits on DM
annihilation cross-section for different DM masses. In giving this limits,
they consider the DM pair annihilate predominantly to $b\bar{b}$ as also
$\tau\bar{\tau}$ and other channels. 

It is possible to assess the total DM content of dSphs galaxies from the
dynamical modeling of the stellar density of the dwarf galaxies and
the velocity dispersion profiles~\citep{Battaglia:2013wqa, Walker:2009zp, Wolf:2009tu}. 
The dynamical masses of
these dwarf galaxies are measured only from stellar velocity 
dispersion and half-light radius. The calculated total mass within 
the half-light radius for a dSphs galaxy is used to 
obtain the integrated $J$-factor of that dSphs galaxy. 
Both the total mass within the half-light radius and the $J$-factor 
are found to be
almost independent on the choice of DM halo profiles~\citep{Martinez:2009jh, Strigari:2013iaa, 
Martinez:2013els}.
Out of the 25 independent dSphs mentioned earlier, $J$-factors of only 18 dSphs are
determined using stellar kinematics data~\citep{Martinez:2013els} while other
seven lack proper statistical significances. 
%Thus, from such
%independent determination of $J$-factors of these dwarf galaxies
%and incorporating uncertainties on these $J$-factors,
Using these uncertainties of $J$-factors, the upper bounds on DM annihilation cross-section for various 
DM masses have been derived with 95\% CL.

We calculate the $\gamma$-ray flux for all the dwarf galaxies.
The $\gamma$-ray spectrum, $\frac{dN}{dE}$ can be obtained for
a given DM mass. The different particle processes for the calculation
of $\frac{dN}{dE}$ are tabulated in Table~\ref{table2}. We compute $\frac{dN}{dE}$
for our benchmark scenario and using the 
integrated $J$-factor
for a particular dSph, the maximum value of the 
velocity-averaged annihilation cross-section ($\langle \sigma v\rangle_{\rm max}$) 
is estimated from the upper bound of the flux
(LHS of Eq.~\ref{gammaflux1}) of that dSph. In this way the upper bounds of 
annihilation cross-section 
%in case of the present inert doublet
%DM candidate, LIP with mass 63.54 GeV 
are computed for all
the 18 dSphs considered and they are tabulated in Table~\ref{table3}. 
%The statistics for the rest of dwarf galaxies is very poor and their $J$-factors
%cannot be determined kinematically. Hence these 7 dSphs are not considered
%in this work. 

\begin{sidewaysfigure}
\begin{center}
%   \centering
%   \includegraphics[angle=-180, width=6in]{dsphs.eps} %[angle=-90, scale=0.75]
\includegraphics[angle=-90, scale=0.75]{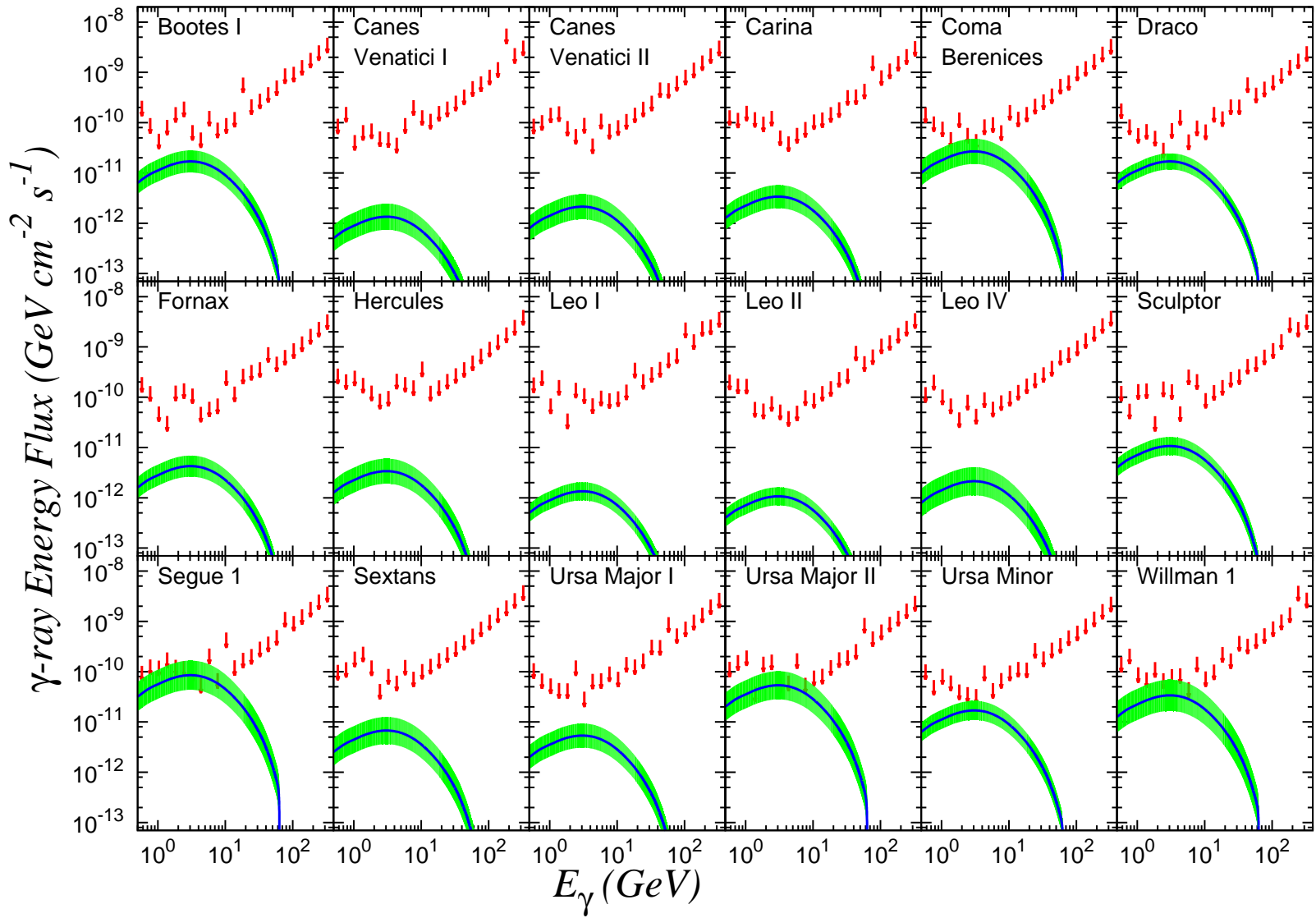}
%\begin{adjustbox}{addcode={\begin{minipage}{\width}}{
   \caption{{\it Comparison of computed $\gamma$-ray flux %   from annihilation DM in IHDM 
   with the bin-by-bin 
   integrated $\gamma$-ray energy-flux upper limits for each dSph. The downward red-coloured
   arrows represent bin-by-bin upper limits on the $\gamma$-ray energy-flux at 95\% CL.
   The blue lines denote the $\gamma$-ray fluxes calculated using the central
   values of integrated $J$-factor whereas the green band is for the uncertainties in the
   measurement of integrated $J$-factors for 18 Milky Way dSphs. See text for details.}}
%    Bootes I, Canes Venatici I, 
%    Canes Venatici II, Carina, Coma Berenices, Draco,
%    Fornax, Hercules, Leo I, Leo II, Leo IV, Sculptor, Segue 1, Sextans, Ursa Major I, 
%    Ursa Major II, Ursa Minor and Willman 1 dsphs. }}
%   \end{minipage}},rotate=90,center}%See text for details.
%   \end{adjustbox}
   \label{dsphs1}
\end{center}
%\end{minipage}
%\end{figure}
\end{sidewaysfigure}

We now compute the $\gamma$-ray flux for all the 18 dSphs considered using Eq.~\ref{gammaflux1}.
The maximum, minimum and central values of integrated 
$J$-factor for each dwarf galaxy which are measured from stellar
kinematics data are tabulated in Table~\ref{table3}~\citep{Ackermann:2013yva}. 
% It is also important to mention that the integrated $J$-factors 
% (obtained by  integrating Eq.~\ref{j_lb_rt} over the solid angle $\Omega$)
% for each dSph galaxy of Table~\ref{table3} are calculated by a 
% line-of-sight integration of squared DM distribution. Then the 
The integration to find $J$-factor involves integration 
over a solid angle $\Delta\Omega$ of
$\sim 2.4\times 10^{-4}$ sr %is performed.
(the field
of view of Fermi-LAT is within the angular radius of $0.5\degr$).
% Note that the field
% of view of Fermi-LAT is within the angular radius of $0.5^{\rm o}$
% which can be translated into a solid angle 
% $\Delta\Omega\sim 2.4\times 10^{-4}$ sr. 
% We have
% calculated the gamma ray flux for the dSph galaxies under 
% consideration considering NFW DM profile for the present DM candidate. 
% Since the value of
% the integrated $J$-factor is almost insensitive to DM halo with 
% factor $\gamma < 1.2$~\citep{Strigari:2007at},
% choice of NFW profile for theoretical calculation does not affect 
% integrated $J$-factor.
%The calculation for the gamma ray spectra
%in this framework for each dwarf galaxy differs from other one
%only by the measured values of integrated $J$-factors. 
The results are shown in the plots given in Fig.~\ref{dsphs1}.%,~\ref{dsphs2} and \ref{dsphs3}.
We also show in Fig.~\ref{dsphs1} %,~\ref{dsphs2} and \ref{dsphs3},
the experimentally obtained bin-by-bin upper limits of the gamma ray energy flux at 95\% CL 
from each dwarf galaxy by downward red-coloured
arrows.
% In Figs.~\ref{dsphs1},~\ref{dsphs2} and \ref{dsphs3}, 
The flux
of a dSph are compared with the upper bound of the flux in each
energy bin. The spread (band) of this flux shown by green indicate the upper
and lower limits of the flux when calculated with the upper
and lower limits of integrated $J$-factors. The photon flux 
calculated using the central value of $J$-factors are shown by blue 
lines in these figures.

%%#######################################################%%
\section{Confronting Extragalactic gamma ray background in this framework\label{EGB_sec}}
%%#######################################################%%

The $\gamma$-rays can also be emitted from DM annihilation in extragalactic
sources and such $\gamma$-rays can be probed for extragalactic DM 
and their origins~\citep{Ullio:2002pj, Bergstrom:2001jj,
Gao:1991rz, Stecker:1978du, Taylor:2002zd, Ng:2013xha,
Ando:2005hr, Oda:2005nv, Pieri:2007ir}.
Such gamma rays from extragalactic sources of DM can remain hidden in the 
huge background of the observationally measured gamma flux by satellite-borne
experiments such as SAS-2 satellite~\citep{1978ApJ...222..833F}, EGRET~\citep{Sreekumar:1997un}, 
Fermi-LAT~\citep{Abdo:2010nz, Ackermann:2014usa}. In order to extract information regarding 
the extragalactic signature of gamma rays from the background one should 
be able to understand and subtract the 
galactic astrophysical components, other sources that may contribute
to the background and the 
backgrounds that the detector may give rise to, in the process of detection.
After this process of subtraction the residual gamma-ray signal thus obtained is found to be
diffuse and isotropic in nature and is known as diffuse isotropic 
gamma-ray background (DIGRB).
Recently in the light of 50-month Fermi-LAT data
an updated tight constraint on DM annihilation 
is given with the modelling of integrated emission of blazars
with such diffuse background absorption~\citep{Ajello:2015mfa}.
This may also be noted that the DIGRB thus obtained embeds in it 
the irreducible contributions from galactic origin as well.
In this section we
% consider the IHDM dark matter (as discussed earlier in this paper) and 
estimate such diffuse isotropic gamma ray background 
or DIGRB for the case of dark matter annihilating into gamma rays
in the framework of the chosen IHDM dark matter candidate.  
We then compare our theoretical calculations with 
EGRET and Fermi-LAT results for extragalactic DIGRB.

\subsection{Formalism} 

\begin{figure}[cbt!]
\begin{center}
\includegraphics[angle=-90, scale=0.4]{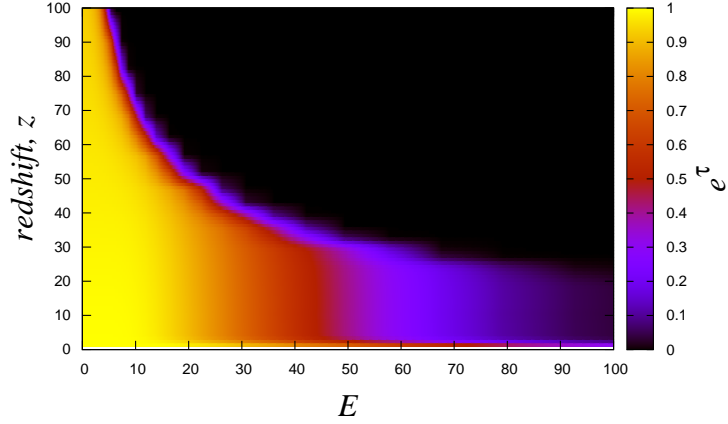}
%%%
   \caption{{\it The variation of optical depth (transparency coefficient)
   $e^{\tau}$ with the energy at detection ($E$) and the redshift ($z$)
   of photon emission for minimum UV background model. The black zone in this figure represents
   total opaque region while the yellow zone is for total transparent one. See text for details.}}
   \label{etau}
\end{center}
\end{figure}

\begin{figure}[cbt!]
\begin{center}
\includegraphics[angle=-90, scale=0.32]{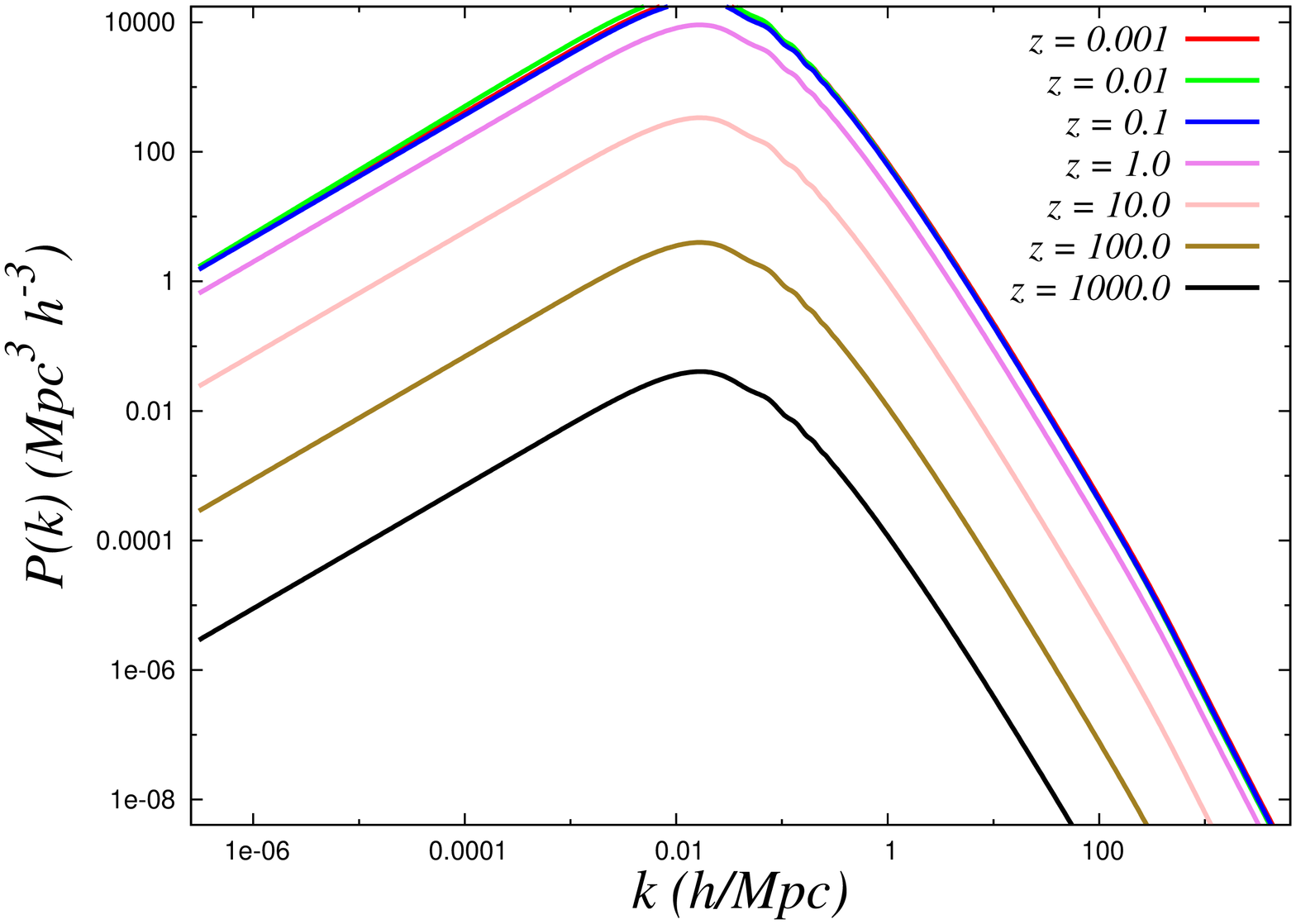}
\includegraphics[angle=-90, scale=0.32]{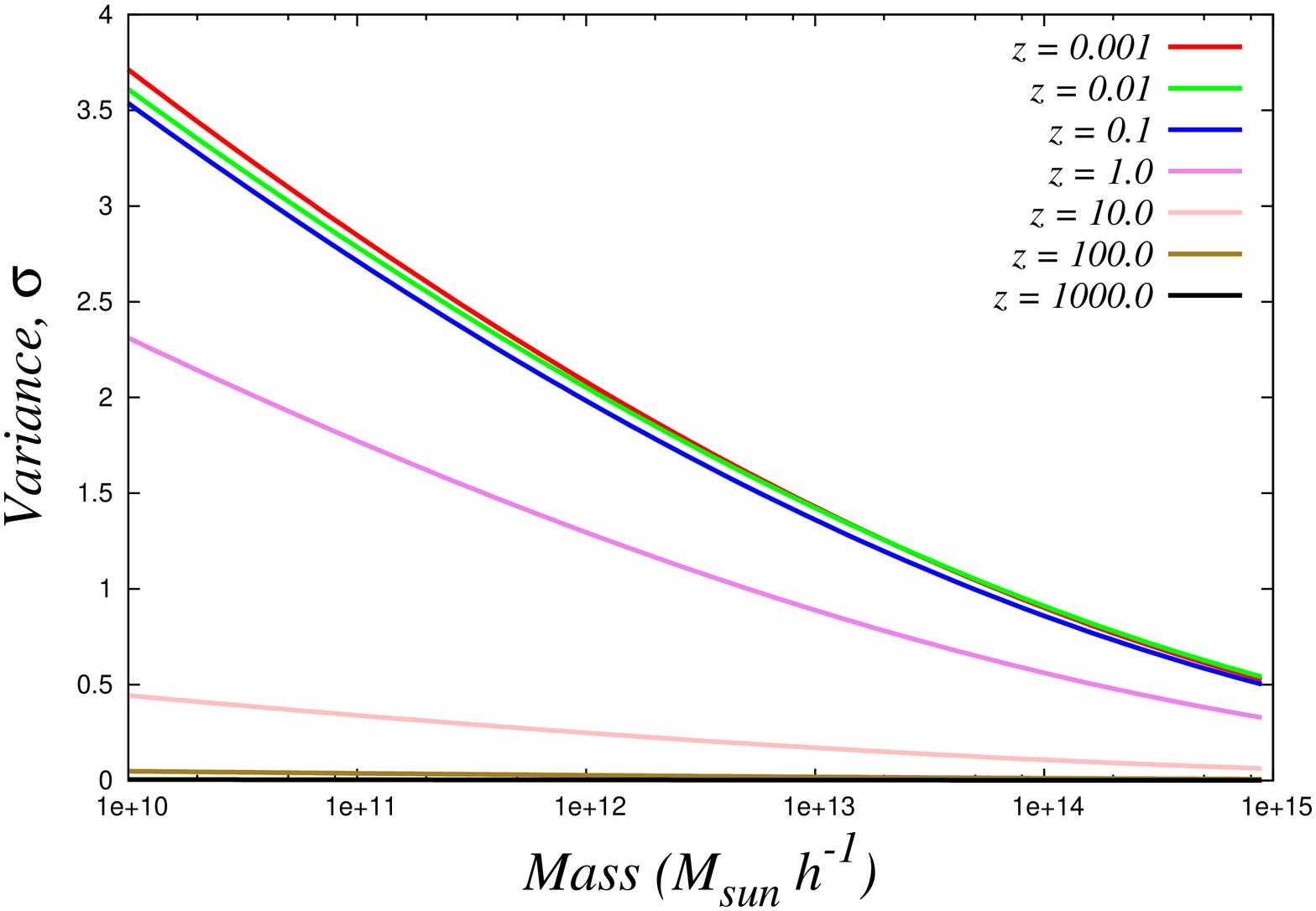}
%%%
   \caption{{\it The variation of the linear power spectrum $P(k)$ of matter 
   density perturbations with the wavenumber $k$ of the fluctuations for different redshifts 
   is shown in the left panel.
   In the right panel the variance $\sigma$ of the density perturbations is shown as 
   a function of halo mass
   for different redshifts. In both plots the values of redshift 
   $z = 10^{-3}, 10^{-2}, 10^{-1}, 10^{0}, 10^{1}, 10^{2} \ \mathrm{and}\ 10^{3}$. 
   See text for details.}}
%    $z = 0.001, 0.01, 0.1, 1.0, 10.0, 100.0 \ \mathrm{and}\ 1000.0$.
   % In the plots $h$ is the usual Hubble constant in the units of 100 km s^{-1} Mpc^{-1}.
   \label{pk_k_var}
\end{center}
\end{figure}

The number of photons which are isotropically emitted from the 
volume element $dV$
in time interval $dt$, in energy range $dE$ and are collected by detector with
effective area $dA$ during time interval $dt_0$ with redshifted energy 
range $dE_0$ can be given by~\citep{Ullio:2002pj},
\beq
  dN_\gamma = e^{-\tau(z,E_0)} \left[(1+z)^3 \, \int dM\;\frac{dn}{dM}(M,z)\,
  \frac{d{\cal N}_{\gamma}}{dE}\left(E,M,z\right)\right]\;
  \frac{dV\, dA}{4\pi [R_0 S_k(r)]^2} \, dE_0 \,dt_0\;.
\label{eq:infflux}
\eeq
In the above, the volume element $dV$ at a  
redshift $z$ is given as 
\beq
  dV = \frac{[R_0 S_k(r)]^2\,R_0}{(1+z)^3} dr d\Omega_{\rm detector}\;,
\label{eq:dv}
\eeq
where $d\Omega_{\rm detector}=\sin\theta d\theta d\phi$ denotes the angular acceptance of the detector. 
In Eq.~\ref{eq:infflux} and Eq.~\ref{eq:dv}, $S_k(r)$ is given by the Robertson 
Walker metric for homogeneous and isotropic universe 
\beq ds^2 = c^2
dt^2 - R^2(t) \left[ dr^2 + S_k^2(r) (d\theta^2 + \sin^2\theta d\phi^2) 
\right]\;, 
\eeq
where $S_k(r)$ signifies the spatial curvature of the universe.
%where $k$ is the curvature parameter and the term $S_k(r)$ is given as
% \beq 
%   S_k(r)=\begin{cases} r,& k=0\;({\rm flat});\cr \arcsin r,& k=+1\;({\rm closed});\cr
%   {\rm arcsinh}\, r,& k=-1\;({\rm open}). 
% \end{cases} 
% \eeq 
%for different curvature of the universe's geometry 
%(flat ($k=0$), closed ($k=+1$) or open ($k=-1$)).
In Eq.~\ref{eq:infflux},
$\frac{d{\cal N}_{\gamma}}{dE}\left(E,M,z\right)$ is the differential photon
energy spectrum for a generic halo with mass $M$ at some redshift $z$. 
The term $\frac{dn}{dM}(M,z)$
is the halo mass function and is defined as number density 
of bound objects with mass $M$
at redshift $z$. The term $e^{-\tau(z,E_0)}$ represents the attenuation of
extragalactic $\gamma$-rays which may come from the absorption of 
these high energy 
$\gamma$-rays on the extragalactic background light (EBL) and
$\tau(z,E_0)$ is the optical depth which is function of $z$ and $E_0$.
The energy and redshift dependence of this attenuation factor is shown in Fig.~\ref{etau}.
Detailed studies regarding this attenuation
are given in Ref.~\citep{Cirelli:2010xx}. 
%The optical depth $\tau(z,E_0)$ 
%represents the nature of absorption at redshift $z$.
%(actually between the redshifts $z'$ and $z''$).
For ultraviolet background we have adopted the minimal model~\citep{Dominguez:2010bv, 
Franceschini:2008tp} obtained after a recent study on blazars. 
%The integral in Eq.~\ref{eq:infflux} is over energy and time ($dtdE$). 
%Note that $dtdE = \frac{dt_0}{(1+z)}.(1+z)dE_0$, 
Note that $dt_0dE_0$ in Eq.~\ref{eq:infflux} is given by
$dtdE = \frac{dt_0}{(1+z)}.(1+z)dE_0$,
where $t_0$ and $E_0$ 
are the time and energy respectively corresponding to the redshift $z =0$.
Summing over all the above contributions, the diffuse extragalactic $\gamma$-ray flux due to DM
annihilation, can be written as,
\bea
  \frac{d\phi_{\gamma}}{dE_0} 
  \equiv \frac{dN_{\gamma}}{dA \,d\Omega \,dt_0 \,dE_0}
  & = & \frac{1}{4 \pi} \int dr \,R_0 e^{-\tau(z,E_0)}
  \int dM\;\frac{dn}{dM}(M,z)\,
  \frac{d{\cal N}_{\gamma}}{dE}\left(E_0\,(1+z),M,z\right) \nonumber \\
  & = & \frac{c}{4 \pi} \int dz \frac{e^{-\tau(z,E_0)}}{H_0\,h(z)}
  \int dM\;\frac{dn}{dM}(M,z)\,
  \frac{d{\cal N}_{\gamma}}{dE}\left(E_0\,(1+z),M,z\right)\;,
\label{eq:flux1}
\eea 
where $c$ is the speed of light in vacuum, $H_0$ is the Hubble constant 
at the present epoch and $h(z)$
%is a redshift dependent function which depends on the choice of the cosmological
%models and can be written in the form,
is written as (for spatially flat universe ($\Omega_k = 0$))
\begin{equation}
  h(z)=\sqrt{\Omega_M(1+z)^3+\Omega_\Lambda}.
\label{eq:hofz}
\end{equation}
where $\Omega_M$ and $\Omega_\Lambda$ are respectively the 
matter and dark energy densities normalised to the critical 
density of the universe.
%and
%$\Omega_k =0$ since the universe is spatially flat.
%Since the observational results indicate the spatial flatness of the 
%universe $\Omega_k =0$. 
%If there is no structure, the above formula reduces to that of 
%the $\gamma$-ray flux 
%produced only from smooth component of dark matter annihilations. 
%In Eq.~\ref{eq:flux1}, the line of sight integral (integral over $dr$) 
%has been replaced by redshift integral (integral over $dz$). 
%\footnote{The relation
%between the co-moving distance, $\chi$ and the redshift $z$ can be written as,
%$\frac{d\chi}{dz} = \frac{c}{H(z)}$, where $H(z) = H_0 h(z)$
%and $\chi = R_0 r$.} 
%Following Press-Schechter~\citep{Press:1973iz}, 
The cosmological dark matter 
halo function $\frac{dn}{dM}(M,z)$ in Eq.~\ref{eq:flux1} can be written
in the form~\citep{Press:1973iz},
\beq 
  \frac{dn}{dM} = \frac{{\rho}_{0,m}}{M^2}
  \nu f(\nu) \frac{d\,\log\nu}{d\,\log M} \;,
  \label{eq:massfunc} 
\eeq
where ${\rho}_{0, m}$ is the comoving background matter density.
%\footnote{${\rho}_{0, m} \simeq \rho_c\Omega_M$ with $\rho_c$ 
%being the critical
%density at the present epoch ($z=0$). More precisely
%$\bar{\rho}_{z, m} = \rho_c\Omega_{\rm M}(1+z)^3$.
%and $M$ is the mass of halo.
%}
In the above, the parameter $\nu = \delta_{\rm sc}/\sigma(M)$, 
defined as the ratio
between the critical overdensity for spherical collapse $\delta_{\rm sc}$ ($\simeq 1.686$) 
and $\sigma(M)$ denotes the variance or
the root mean square density fluctuations of the linear density field in sphere
that contains the mean mass $M$. The term $\sigma^2 (M)$ can be written 
in terms of the linear power spectrum $P(k)$ of the fluctuations~\citep{Sheth:1999mn} as,
\beq 
  \sigma^2(M) \equiv \int d^3k \;
  \tilde{W}^2(kR) \, P(k) \,\,\, ,
\eeq  
where $\tilde{W}(kR)$ is the Fourier transform of the top hat window function
and $R$ is the comoving length scale. For collapsed halos, the mass is found to
be in the form $M \simeq (4/3)\pi R^3 \rho_c (z_c)$ with $z_c$ being the redshift where 
collapsing of halos occurs. 
The power spectrum $P(k)$
can be parametrised as $P(k) \propto k^n T^2(k)$ where $n$ is the spectral index and
$T$ is the transfer function that incorporates the effect of scale dependency 
of the primordial power spectrum generated during inflation. This transfer function
depends on the nature of DM and baryon density in the universe.
Thus the transfer function can be calculated from the cosmic microwave background
data. %considering using \texttt{CAMB} code.
The variation of the power spectrum $P(k)$ with wavenumber $k$ for different
redshifts is shown in the left panel of Fig.~\ref{pk_k_var}. The right panel of
Fig.~\ref{pk_k_var} corresponds to the plot showing the variation of variance $\sigma$
with halo mass $M$ for different values of redshift.
The function $f(\nu)$ in Eq.~\ref{eq:massfunc}, known as the multiplicity 
function,
can be modelled in the ellipsoidal collapse model~\citep{Sheth:1999mn} by,
\beq
  \nu f(\nu) = 2 A \left(1+\frac{1}{\nu'^{2p}}\right)
  \left(\frac{\nu'^{2}}{2\pi}\right)^{1/2}
  \exp\left(-\frac{\nu'^{2}}{2}\right)\,\,\, ,
  \label{eq:nuofnu}
\eeq
where $\nu'=\sqrt{a}\nu, a = 0.707, p = 0.3$ are obtained by 
fitting Eq.~\ref{eq:massfunc}
to $N$-body simulation of Virgo consortium~\citep{Jenkins:1997en}. 
%The value of the parameter $A$ can be
%fixed by using the relation $\int d\nu f(\nu) = 1$ which follows from
%the condition that the total mass should lie within a given halo, i.e, 
%$\int dM \;M dn/dM = {\rho}_0$.
The value of $A$ is obtained to be $0.3222$
%This fixes the value of $A$ to be $0.3222$.
For the choice of parameter values, 
$a=1, p=0$ and $A = 0.5$, Eq.~\ref{eq:massfunc} reduces to the original
Press-Schechter theory~\citep{Press:1973iz}. It is found in $N$-body
simulations that the estimations of higher and lower mass halos differ 
from that predicted by Press-Schechter model. This problem can be handled
in Sheth-Torman model
by considering ellipsoidal collapse model instead of spherical one.

\begin{figure}[cbt!]
\begin{center}
\includegraphics[angle=-90, scale=0.32]{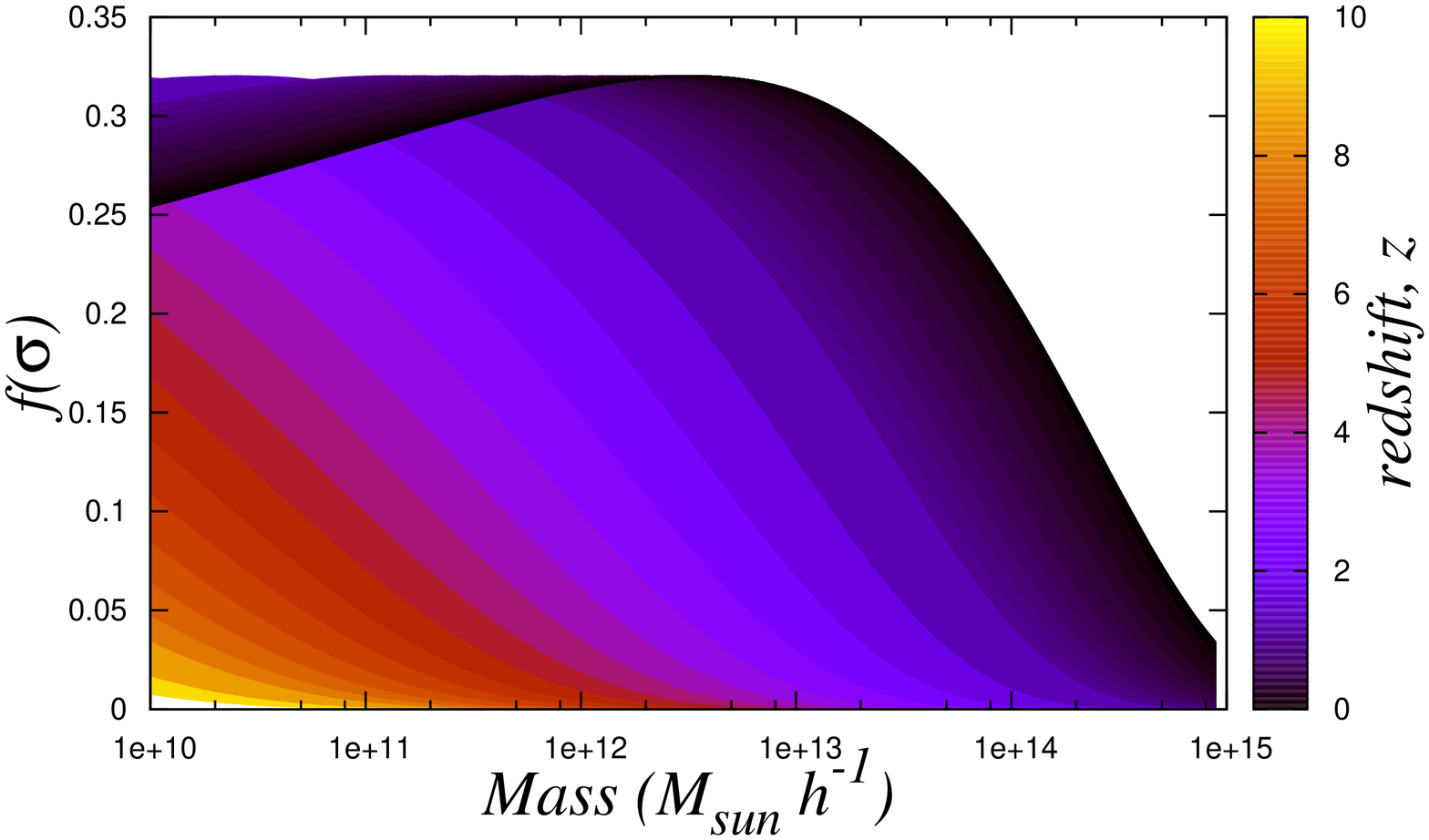}
\includegraphics[angle=-90, scale=0.32]{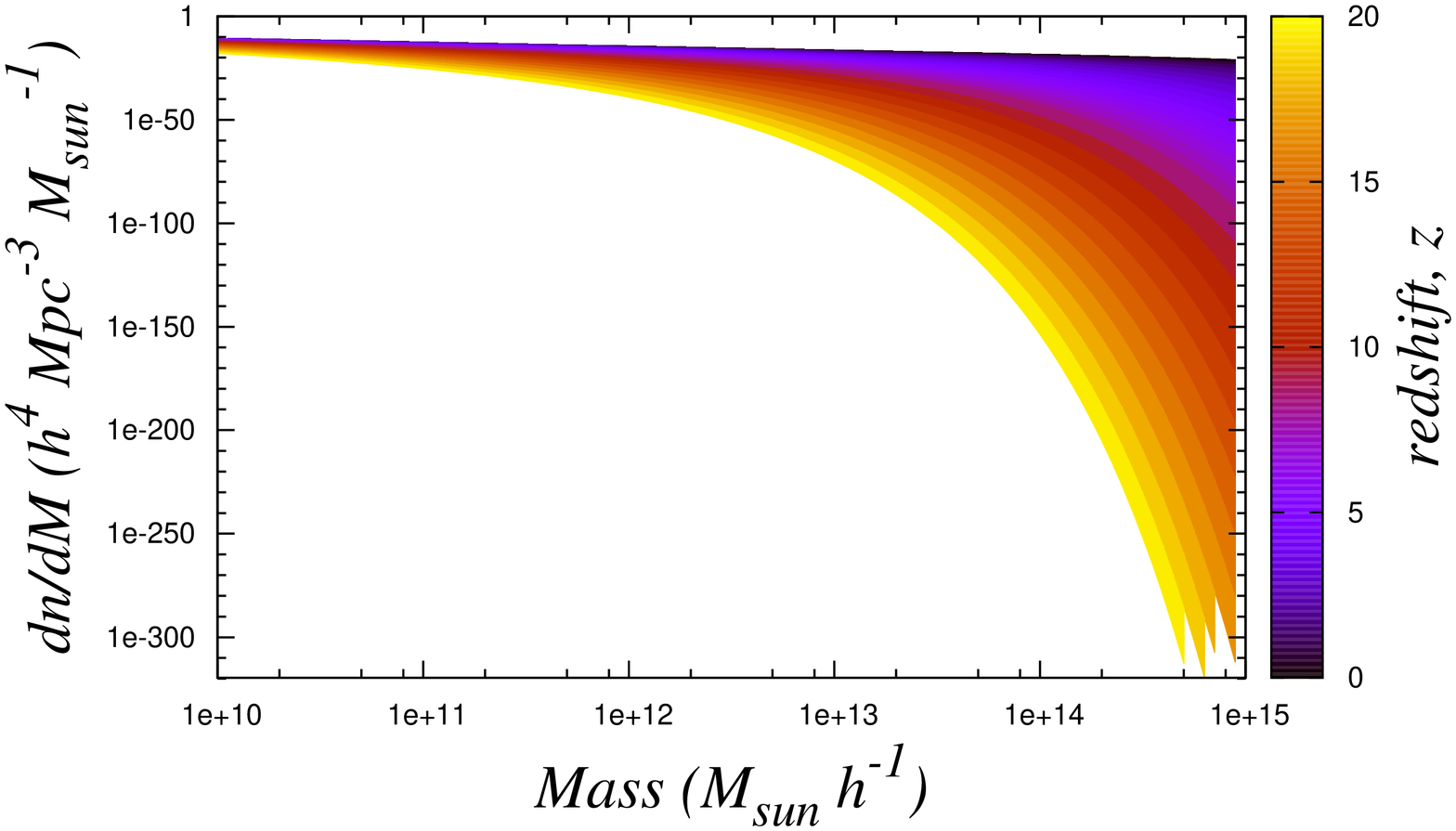}
%%%
   \caption{{\it The fraction of mass collapsed, $f(\sigma)$ in Sheth-Torman model for 
   different redshifts $z$ and the halo masses $M$ is shown in left panel. 
   The variation of Sheth-Torman halo mass function ${dn\over dM}$ with 
   the redshift $z$ and the halo mass $M$ is shown in right panel.
   See text for details.}}
   \label{z_m_fsig_dndm}
\end{center}
\end{figure}

In the left panel of Fig.~\ref{z_m_fsig_dndm} the variations of the fraction of mass collapsed or 
$f(\sigma)$ in the ellipsoidal collapse model with redshifts $z$ and 
the halo mass $M$ are shown. Note that $f(\sigma)$ as shown in the left panel of 
Fig.~\ref{z_m_fsig_dndm} can be obtained 
by simple transformation of $f(\nu)$ by plugging in $\nu = \delta_{\rm sc}/\sigma(M)$ and is given by 
$f(\sigma) =A\sqrt{\frac{2a}{\pi}}\left[1+\left(\frac{\sigma^2}{a\delta_{\rm sc}^2}\right)^p\right]
\frac{\delta_{\rm sc}}{\sigma}\exp\left[-\frac{a\delta_{\rm sc}^2}{2\sigma^2}\right]$.
In the right panel of Fig.~\ref{z_m_fsig_dndm} we have shown the 
variations of the considered halo mass function $dn/dM$ of Sheth-Torman model~\citep{Sheth:1999mn}
with redshift $z$ as well as with the halo mass $M$.
%The numerical calculations for these plots
All the numerical calculations related to Fig.~\ref{z_m_fsig_dndm} %,~\ref{z_m_dndm} 
have been performed using \texttt{HMFcalc}~\citep{Murray:2013qza} code.

For the halo profile we have chosen NFW halo profile~\citep{Navarro:1995iw, Navarro:1996gj} given by,
\beq 
  \rho(r)=\rho_s \,g(r/r_s)=\rho_s {1\over {r\over r_s} \left(1+{r\over r_s}\right)^2}\;,
  \label{eq:haloprof} 
\eeq   
Any DM halo of mass $M_h$ enclosed at a radius $r_h$ is,
\beq
M_h = 4\pi\rho_sr_h^3f(r_s/r_h)\,\,\, , 
\eeq
where $f(x) = x^3[\ln(1+x^{-1}) - (1+x)^{-1}]$.

Also a halo of mass $M$ at some redshift $z$ can be written in terms of
mean background $\bar{\rho}(z)$ as,
\beq
M \equiv  {4\pi\over 3} \Delta_{vir} \bar{\rho}(z)\, R_{vir}^3\,\,\, ,
\eeq
where $R_{vir}$ is the virial radius defined as the radius within which
the total halo mass $M$ is contained with mean halo density $\Delta_{vir} \bar{\rho}(z)$.
The term $\Delta_{vir}$ is the virial overdensity
with respect to the mean matter density which may depend on the cosmological parameters 
but independent of halo mass $M_h$. For the flat cosmology, $\Delta_{vir}(z)$ 
can be cast into the following form~\citep{Bryan:1997dn},
\beq
\Delta_{vir} \simeq (18\pi^2 + 82d
- 39 d^2)\,\,\, ,
\eeq
with $d\equiv d(z) = {\Omega_m(1+z)^3 \over (\Omega_m(1+z)^3 + \Omega_{\Lambda})} - 1$. We choose
the value of $\Delta_{vir}(z)$.%% to be 200. %%for Einstein-D'Sitter spacetime~\citep{}.

The $\gamma$-ray energy spectrum $\frac{d{\cal N}_{\gamma}}{dE}\left(E_0\,(1+z),M,z\right)$ (Eq.~\ref{eq:flux1})
%which is the differential gamma-ray energy spectrum 
for the gamma-ray emitted inside a halo of mass $M$ 
at redshift $z$ is written to the form,
\beq
  \frac{d{\cal N}_{\gamma}}{dE} (E,M,z) = \frac{\langle\sigma v\rangle}{2}
  \frac{dN_{\gamma}(E)}{dE} \int dc^{\,\prime}_{vir}\; 
  {\cal{P}}(c^{\,\prime}_{vir})
  \left(\frac{\rho^{\prime}}{M_{\chi}}\right)^2 
  \int d^3r\; g^2(r/a)\; ,
\label{eq:dnde}
\eeq
where $\langle\sigma v\rangle$ is the thermally averaged value of annihilation
cross-section times the relative velocity, $\frac{dN_{\gamma}(E)}{dE}$ is the differential
$\gamma$-ray energy spectrum produced per unit annihilation of dark matter and $M_{\chi}$
is the mass of dark matter. The log-normal 
distribution ${\cal{P}}(c_{vir})$ of the 
concentration parameter $c_{vir}$ around the mean value is
chosen within $1\sigma$ deviation~\citep{Sheth:2004vb}, for halos with mass $M$. 
Finally one can write,
\beq
  \frac{d{\cal N}_{\gamma}}{dE} (E,M,z) = \frac{\sigma v}{2} \frac{dN_{\gamma}(E)}{dE}
  \frac{M}{M_{\chi}^2} \, \frac{\Delta_{vir}\bar{\rho} (z)}{3}\,
  \int dc^{\,\prime}_{vir}\; {\cal{P}}(c^{\,\prime}_{vir})
  \frac{(c^{\,\prime}_{vir}\,r_{-2})^3}
  {\left[I_1(c^{\,\prime}_{vir}\,r_{-2})\right]^2} \,
  I_2(x_{min},c^{\,\prime}_{vir}\,r_{-2})\; .
\label{eq:dnde2}
\eeq
In the above $r_{-2}$ is the ratio between $r_s^{(-2)}$ and $r_s$
where $r_s^{(-2)}$ is the radius at which the effective logarithmic slope $-2$
that follows from the relation, 
$d/dr\left.\left(r^2 g(r)\right)\right|_{r=r_s^{(-2)}} = 0$.
For NFW profile, $r_s^{(-2)} = r_s$.
Hence $c_{vir}r_{-2} = R_{vir}/r$ and 
the form of integration $I_n(x_{min},x_{max})$
in Eq.~\ref{eq:dnde2} can be cast into the form,
\beq
  I_n(x_{min},x_{max}) = \int_{x_{min}}^{x_{max}} dx\, x^2 g^n(x)\;.
\eeq
Plugging the above equation in Eq.~\ref{eq:flux1}, the analytic form of extragalactic
gamma-ray flux from DM annihilation can be obtained as~\citep{Ullio:2002pj}
\beq
  \frac{d\phi_{\gamma}}{dE_0} = \frac{\sigma v}{8 \pi} \frac{c}{H_0}
  \frac{{\rho}_0^2}{M_{\chi}^2} \int dz\;(1+z)^3 \frac{\Delta^2(z)}{h(z)}
  \frac{dN_{\gamma}(E_0\,(1+z))}{dE} e^{-\tau(z,E_0)} \;,
\label{eq:flux2}
\eeq
where the expression for $\Delta^2(z)$ can be given by,
\beq
  \Delta^2(z) \equiv \int dM \frac{\nu(z,M) f\left(\nu(z,M)\right)}{\sigma(M)}
  \left|\frac{d\sigma}{dM}\right| \Delta_M^2(z,M)\; 
\label{eq:D2}
\eeq
with 
\beq
  \Delta_M^2(z,M) \equiv
  \frac{\Delta_{vir}(z)}{3}\,\int dc^{\,\prime}_{vir}\; 
  {\cal{P}}(c^{\,\prime}_{vir})
  \frac{I_2(x_{min},c^{\,\prime}_{vir}(z,M)\,r_{-2})}
  {\left[I_1(x_{min},c^{\,\prime}_{vir}(z,M)\,r_{-2})\right]^2} 
  (c^{\,\prime}_{vir}(z,M)\,r_{-2})^3 \;.
\label{eq:D2M}
\eeq

In all of the above the concentration parameter, $c_{vir}$ is defined as
\beq
  c_{vir} = \frac{R_{vir}}{r_s^{(-2)}}\,\,\, ,
\eeq 
We have chosen two forms of concentration parameter $c_{vir}$ following 
Macci\`{o} {\it et al.}~\citep{Maccio':2008xb} and 
power law model~\citep{Neto:2007vq, Maccio':2008xb}.
For the first choice (Macci\`{o} {\it et al.}),  
$c_{vir}(M,z)=k_{200} 
\left(\mathcal{H}(z_f(M))/\mathcal{H}(z)\right)^{2/3}$, 
where $k_{200} \simeq 3.9$, $\mathcal{H}(z)=H(z)/H_0$ and 
$z_c(M)$ is the effective redshift during the formation of a halo with mass $M$.
In the power law model (second choice) however the expression of $c_{vir}(M,z)$
is adopted as 
$c_{vir}(M,z)=6.5\, \mathcal{H}(z)^{-2/3}$ $(M/M_*)^{-0.1}$, 
$M_*=3.37\, 10^{12} h^{-1}M_\odot$. 
This choice of $c_{vir}(M,z)$ provides
a reasonable fit within the resolved mass range in the simulations.

The dark matter substructures are present within halo and form bound objects.
The mass of the smallest possible such bound object (subhalo) is denoted as $M_{\rm min}$.
The value of this minimum subhalo mass, $M_{\rm min}$ is %can be 
determined from the temperature at which the
DM particles just start decoupling kinematically
from the cosmic background.% particles in the universe.

In this work we perform our analysis for two typical values of $M_{\rm min}$; %minimum masses of these subhalos 
$M_{\rm min} =10^{-6} M_{\odot}$ and
$10^{-9}\, M_{\odot}$ \citep{Martinez:2009jh, Bringmann:2009vf}.
The boost factor for $\gamma$-ray flux due to these subhalos
depends inversely on $M_{\rm min}$.

\subsection{Non-DM Contributions in DIGRB}
The extragalactic gamma-ray spectrum in the energy range 
between $\sim$ few hundred MeV and $\sim$ few hundred GeV 
as observed by the Fermi-LAT telescope is found to follow 
almost a power law spectrum (${dN_{\gamma}\over dE} \propto E^{-2.41}$). 
There are contributions from astrophysical sources other than that from possible dark matter
annihilation~\citep{Tavakoli:2013zva}. 
The possible sources that contribute to the diffuse $\gamma$-ray background
other than the DM include 
BL Lacertea objects (BL Lacs), flat spectrum radio quasars (FSRQs), millisecond pulsars (MSPs),
star forming galaxy (SFG), Fanarof-Riley (FR) radio galaxies of type I (FRI) and type II (FRII),
ultra high energy cosmic rays (UHECRs), gamma ray bursts (GRBs), star burst galaxy (SBG), 
Ultra High Energy protons in the inter-cluster material (UHEp ICM) and 
gravitationally induced shock waves (IGS) etc.
The spectral features of photon spectra originated from the non-DM objects~\citep{Tavakoli:2013zva}
considered in this work 
are concisely summarised in Table~\ref{table4}. 
%All of these minimal non-DM contributions discussed above, in principle, be
%summed up and it is found that they make up for $\sim 40\%$ of 
%the extragalactic gamma-ray 
%observed by the Fermi-LAT telescope~\citep{Abdo:2010nz}.

\begin{figure}[cbt!]
\begin{center}
\includegraphics[angle=-90, scale=0.32]{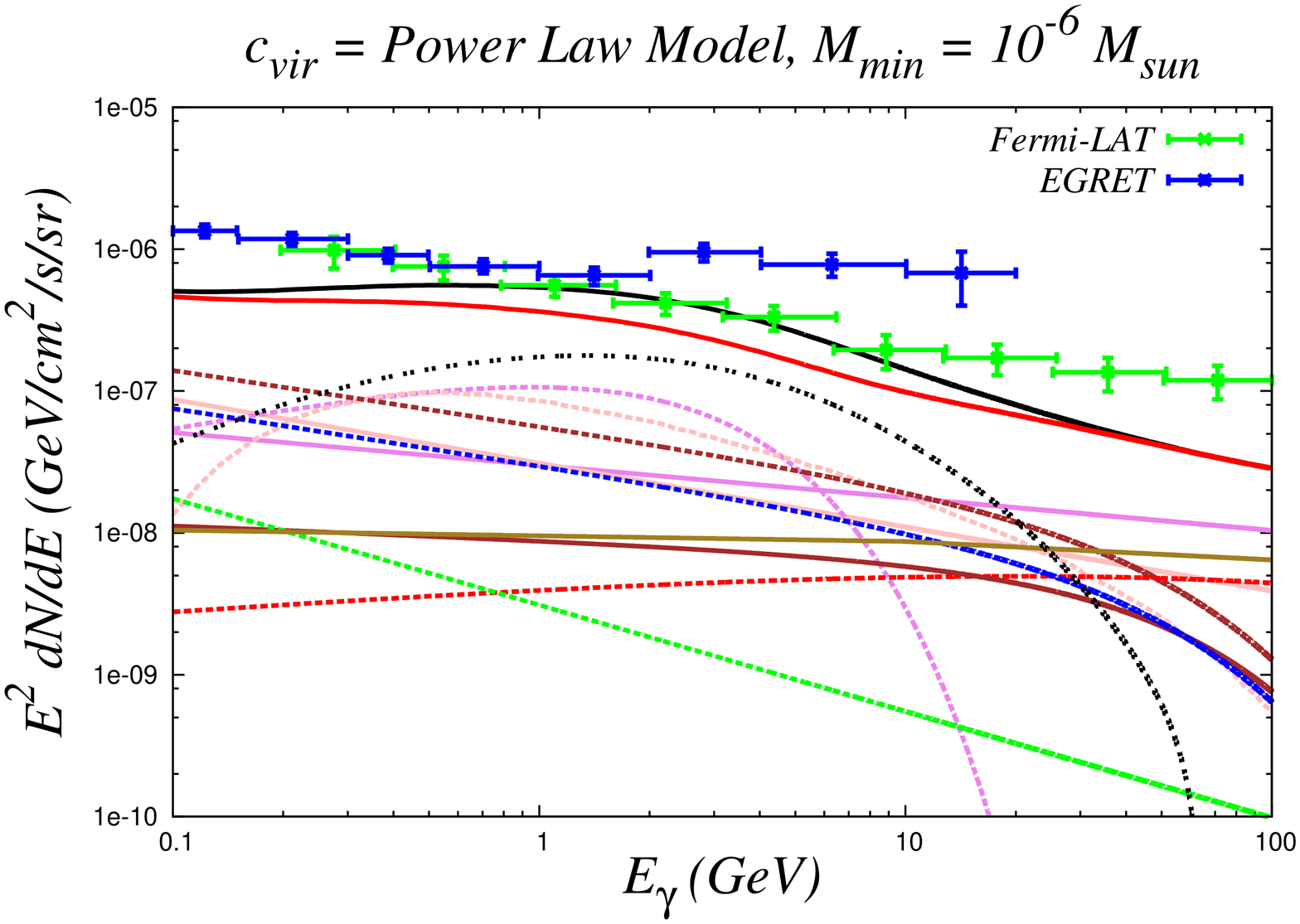}
\includegraphics[angle=-90, scale=0.32]{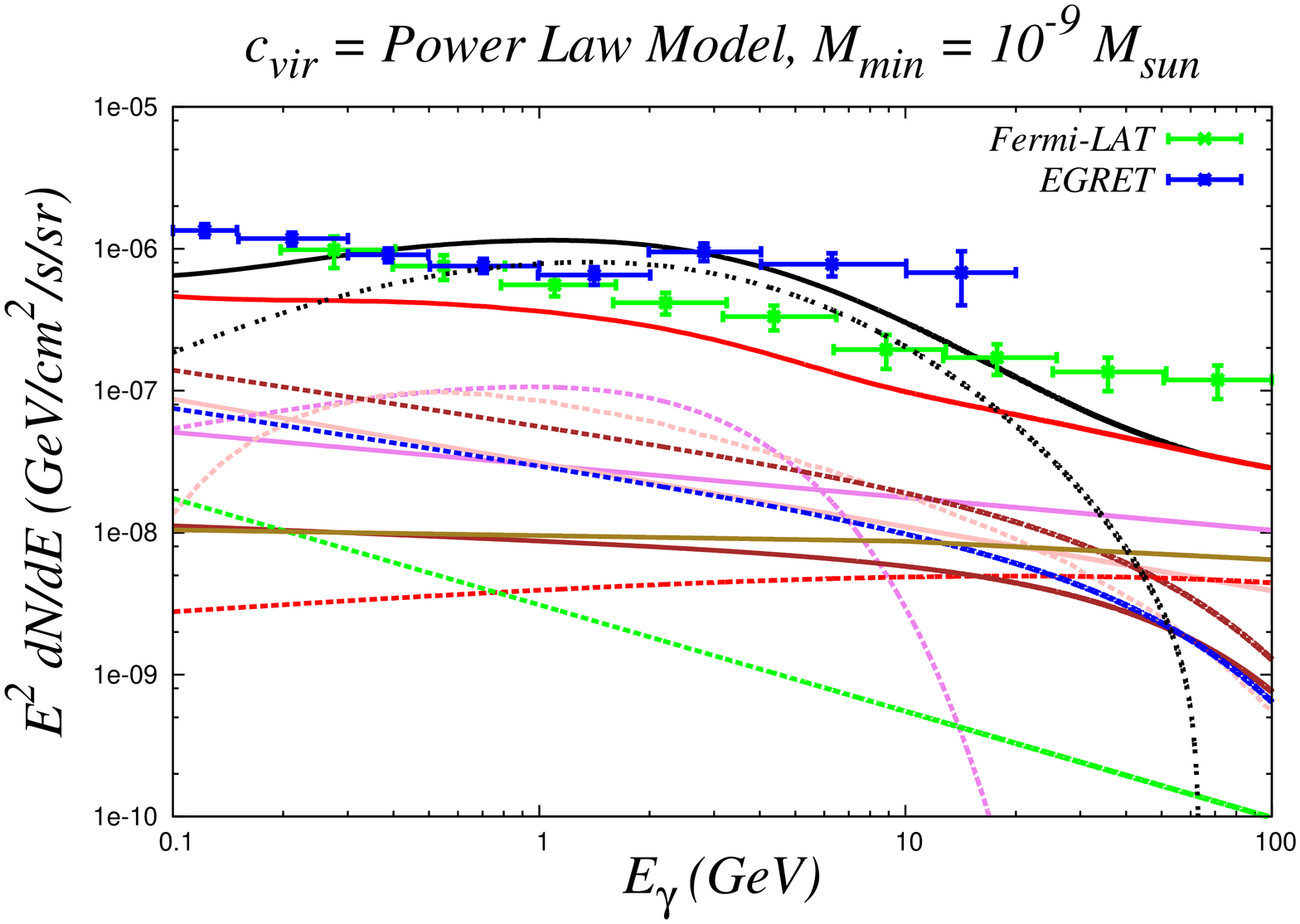}
\includegraphics[angle=-90, scale=0.32]{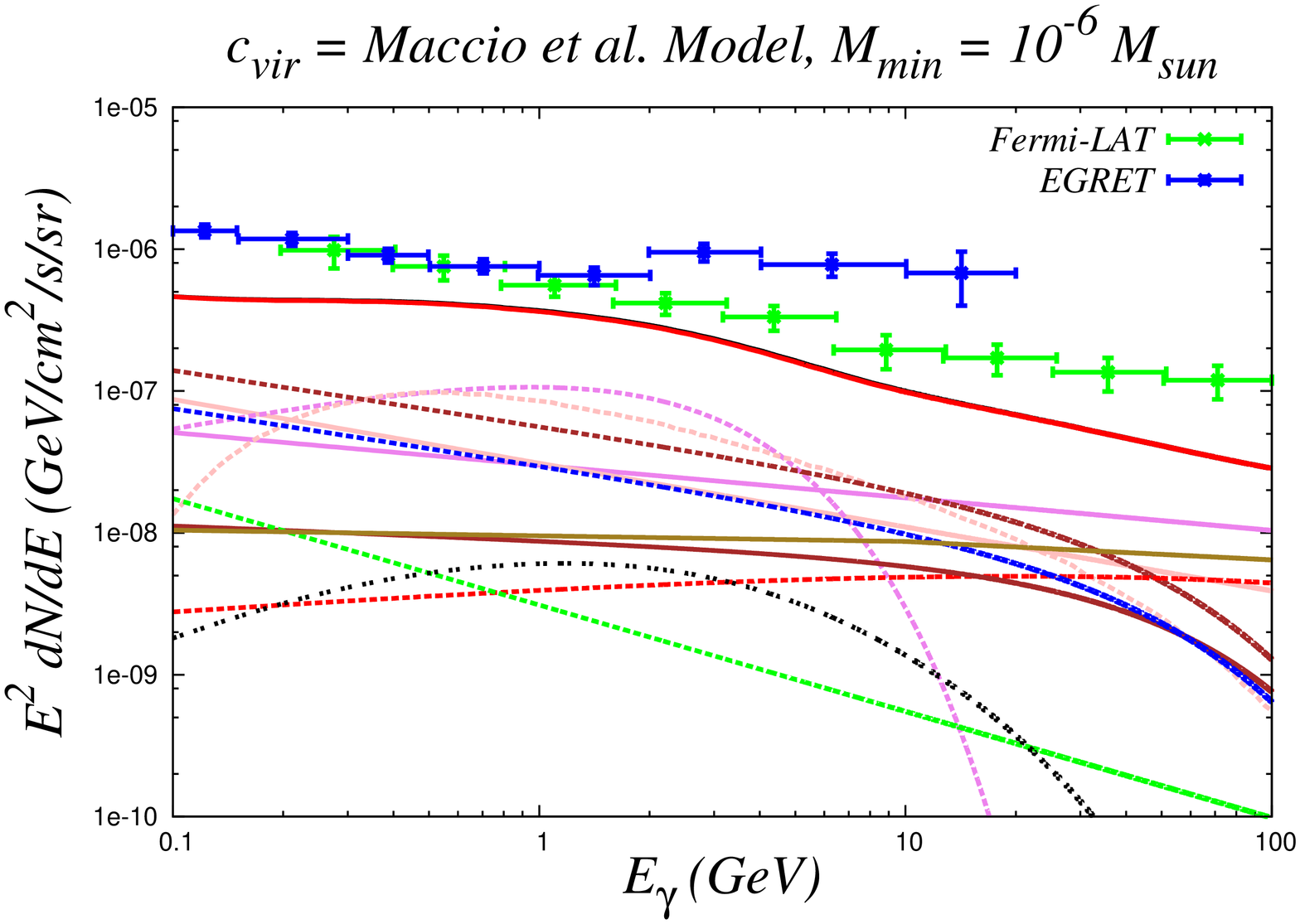}
\includegraphics[angle=-90, scale=0.32]{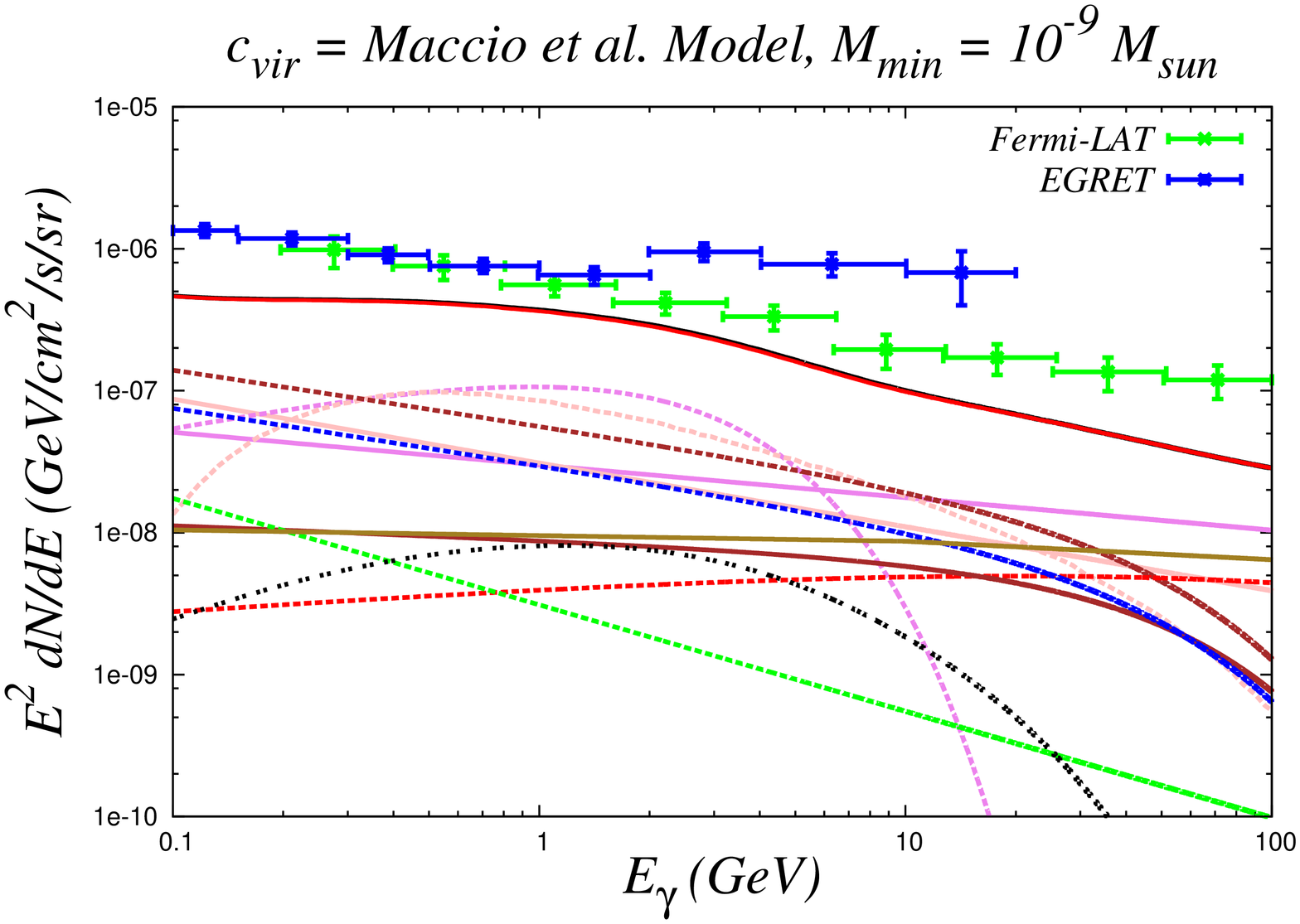}
\includegraphics[angle=-90, scale=0.32]{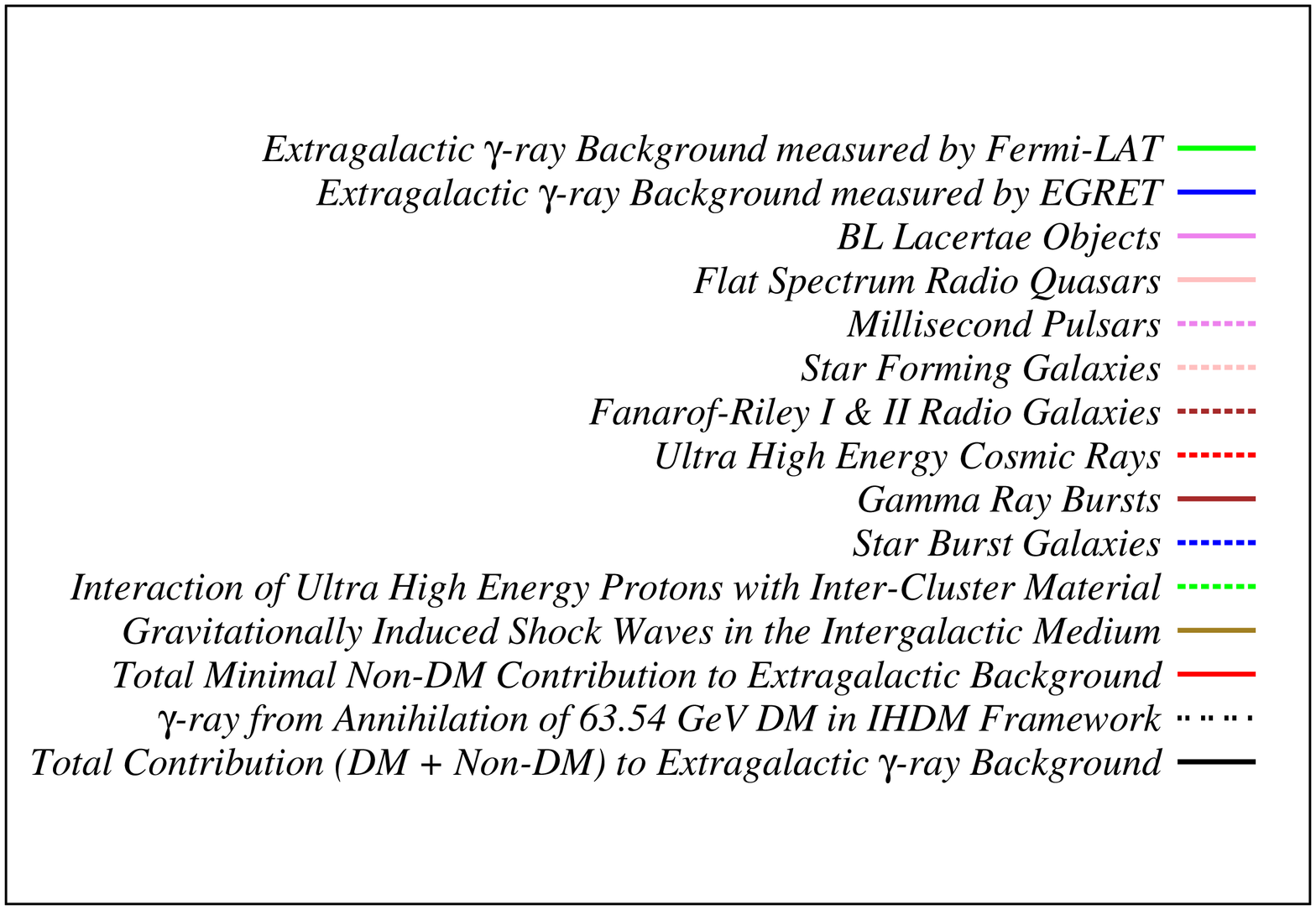}
%%%
   \caption{{\it The observed extragalactic $\gamma$-ray fluxes by EGRET and Fermi-LAT 
   are compared with the 
   sum total value of the $\gamma$-ray fluxes obtained from the present calculation.
   The calculated value of extragalactic $\gamma$-ray flux is obtained by summing over 
   the $\gamma$-ray flux calculated from DM annihilation for IHDM LIP DM
   (considered in this work) and other possible $\gamma$-rays (extragalactic $\gamma$-ray
   sources of non-DM origin) from extragalactic sources. See text for details.}}
%    Comparison of the sum total values of extragalactic gamma-ray fluxes 
%    composed of the contributions both from dark matter and 
%    minimal non dark matter with the observations from EGRET and Fermi-LAT}}
   \label{total_exgal}
\end{center}
\end{figure}

We add up the contributions to EGB both from the annihilating DM  
in IHDM framework ($\sim$ 63.5 GeV DM considered in this work) and 
the other possible non-DM astrophysical
sources. The comparison of the sum total value of the $\gamma$-ray flux with the observed 
EGB by EGRET and Fermi-LAT is shown in Fig.~\ref{total_exgal}. Needless to mention here that
the four plots in Fig.~\ref{total_exgal} are for different parametrisations of concentration
parameter $c_{vir}$ and subhalo mass $M_{\rm min}$. As mentioned earlier we have considered
BL Lacs, FSRQs, MSPs, SFGs, FR (type I and II), UHECR, GRBs, SBGs, UHEp interacting with ICM 
and IGS as contributors to EGB other than DM and their contributions to EGB are shown with different
lines in Fig.~\ref{total_exgal}. The computed total photon spectra 
in the plots of Fig.~\ref{total_exgal} are shown
by black solid lines while the red solid line is for the minimal non-DM contribution to EGB. 
The black lines are found to be on top of the red lines for Macci\`{o} {\it et al.} models.

\begin{figure}[cbt!]
\begin{center}
\includegraphics[angle=-90, scale=0.4]{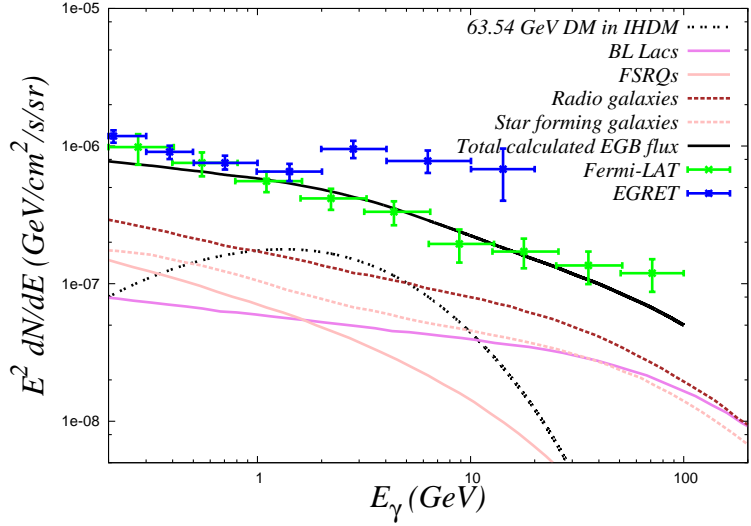}
%%%
   \caption{{\it Comparison of the observed 
   $\gamma$-ray fluxes by EGRET and Fermi LAT
   theoretical results for $\gamma$-ray spectra for $\sim$ 63.5 GeV DM in IHDM considering the 
   modelling of extragalactic and astrophysical parameters as done in Ref.~\cite{Cholis:2013ena}. 
   See text for details}}
   \label{dan_exgal}
\end{center}
\end{figure}

We also mention that the extragalactic $\gamma$-ray signal is analysed within the dark matter
annihilation scenario in Ref.~\citep{Cholis:2013ena}. 
In their analysis they have constructed a model for
the non-DM astrophysical contributions to the extragalactic $\gamma$-ray background and they
have also adopted a substructure model based on numerical simulations. 
The authors in Ref.~\citep{Cholis:2013ena} have considered 
subhalo boost factor $b_{sh}$ in their analyses to be the following form~\citep{Ando:2013ff}
\beq
b_{sh}(M) \approx 110\times(M_{200}/10^{12}M_{sun})^{0.39}
\eeq
where $M_{200}$ denotes the mass enclosed within a radial region where 
the avaraged density is 200 times more than the critical density of the universe.
We have performed the
calculation of extragalactic photon flux for DM for the best-fit model parameter in IHDM 
based on their extragalactic modelling. The result is shown in Fig.~\ref{dan_exgal}. In 
Fig.~\ref{dan_exgal} we have only considered the contributions to EGB 
only from radio galaxies, BL lacs, FSRQs and SFGs other than that from DM annihilation. The sum total 
contributions to EGB is found to fit reasonably well with Fermi-LAT data.

\subsection{Galactic (sub)halo contribution to DIGRB}

There could be a significant contribution to DIGRB which is of galactic origin
along the line of sight due to the passing of the signal from extragalactic sources through the Milky
Way galactic halos and subhalos.
% apart from that coming from the extragalactic DM signal and from non-DM extragalactic signals. 
%Different N-body simulations predict highly galactocentric smooth 
%DM density profiles far beyond our visible galaxy. Also 
From numerical simulation the main DM
halo is found to host a large amount of substructures in form of subhalos
\citep{Springel:2008zz,Diemand:2007qr}. 

%The DM density profile of galactic main halo, in principle, yields an anisotropic 
%$\gamma$-ray signal from DM annihilation. But
The signal from the DM annihilation 
at the galactic substructures could potentially give rise to an almost isotropic signal since 
this generated $\gamma$-ray flux is proportional to the 
less centrally concentrated number density distribution of subhalos 
%which is not the case of
%the smooth DM annihilation signal in the main halo DM distribution 
\citep{Springel:2008cc}.
%The smooth galactic halo assumes only the host halo density without any effect of 
%substructures embedded in it.
The averaged photon intensity from DM annihilation 
in such smooth halo of the Milky Way can be written as,
\beq 
\frac{dI_{\rm sm}(E_{\gamma})}{dE_{\gamma}} 
= \frac{\langle \sigma v \rangle}{2 m_{\rm DM}^2} \frac{dN_\gamma}{dE_{\gamma}}  
\frac1{\Omega_e}  \int_{V_*} dV ~ \frac{ \rho_{\rm MW}^2(s,b,\ell)}{4\pi s^2},
\label{smoothfinal}
\eeq
where $s$ and $\Omega_e$ are the distance from the galactic centre and the observed solid angle.
In the above $b$ and $\ell$ are galactic coordinates (latitude and longitude respectively) chosen
to be $30^\circ\leq |b| \leq 90^{\circ}$, $0\leq \ell <2\pi$~\citep{Cholis:2013ena}. 
$r_{s, {\rm MW}}=21.5$ kpc, $r_{\rm vir, MW}=258$ kpc, 
$\rho_{s, {\rm MW}}=4.9\times10^6\, M_{\odot}$ kpc$^{-3}$,
$M_{\rm vir, MW}=1.0\times 10^{12}\, M_{\odot}$~\citep{Klypin:2001xu} 
are chosen for our calculation.~\footnote{The NFW halo profile of Milky Way is chosen 
as $\rho_{\rm MW}(r) = {\rho_{s, {\rm MW}}\over(r/r_{s,{\rm MW}})(1+r/r_{s,{\rm MW}})}$
where $r$ and $r_{s,{\rm MW}}$ are the galactocentric and scale radii respectively and 
$\rho_{s, {\rm MW}}$ is the scale density. The above-mentioned halo profile 
is assumed to extend up to the virial radius $r_{\rm vir, MW}$ with the virial mass $M_{\rm vir, MW}$.}

The photon flux produced in the smooth halo component are much subdominant than that yielded
in the subhalos and hence they contribute negligibly to extragalactic $\gamma$-ray background.
In $\Lambda$CDM cosmology, the formation of the structures is assumed to be hierarchical.
The smaller DM halos are formed first and the larger ones later. In the period
of structure formation the smaller halos are tidally disrupted after being
captured by the larger host halos of galaxies and clusters 
and hence the outer low density layers are stripped in this process. Thus
the central dense cores only survive and behave as subhalos of the host
halos. These substructures of DM halo enrich DM phenomenology by giving rise to  
substantially enhancement of the DM annihilation rates within a halo.
The contribution to differential gamma-ray flux from subhalo can be obtained from the 
differential luminosity profile of each subhalo which is given by,  
\beq
\frac{dL_{\gamma}}{dE_{\gamma}}=\frac{\langle \sigma v \rangle}{2 m_{\rm DM}^2} 
\frac{dN_\gamma}{dE_{\gamma}} \int dV_{\rm sh}\, \rho_{\rm sh}^2.
\label{sat:lum}
\eeq
For an individual subhalo with mass $M$, the photon intensity can be written as,
\begin{eqnarray}
\frac{d{\mathcal I}(E_{\gamma},s,M)}{dE_{\gamma}}= \frac1{4\pi s^2} \frac{d L(E_{\gamma},M)}{dE_{\gamma}}
=\frac1{4\pi s^2} \frac{b_{\rm gs} \langle \sigma v \rangle}{2 m_{\rm DM}^2} \frac{dN_\gamma}{dE_{\gamma}} \frac{M^2}{r_{s,{\rm sh}}(M)^3} {\mathcal G}[c_{\rm cut}(M)] .
\label{sat:individual}
\end{eqnarray}
where $r_{s,{\rm sh}}$ denotes the scale radius of the subhalo. In the above
the factor $b_{\rm gs}$ determines the contribution from substructure within 
each subhalo (`subsubhalo') and is chosen to be 2~\citep{Kuhlen:2008aw}.
The function ${\mathcal G}[c_{\rm cut}(M)]$ which can be obtained using integral over the volume of each satellite
and the form of subhalo concentration $c_{\rm cut}$~\footnote{
We choose the DM density within each subhalo of mass $M$ to be truncated NFW halo profile,  
\begin{equation}
  \rho_{\rm sh}(r_{\rm sh}|M) =
  \left\{\begin{array}{lcc}
  \rho_{\rm NFW}(r_{\rm sh}|M) & \mbox{for} & r_{\rm sh}
  \le r_{\rm cut}, \\
  0 & \mbox{for} & r_{\rm sh} > r_{\rm cut}, \end{array} \right.
  \label{eq:truncated NFW}
\end{equation}
where $c_{\rm cut}$ is the cutoff radius for this profile.
} following Ref.~\citep{Ando:2009fp}, can be given as,
\beq
{\mathcal G}[c_{\rm cut}(M)]=\frac{1}{12\pi} \left[ 1 - \frac{1}{\left( 1 + c_{\rm cut}\right)^3}\right] \left[ \ln(1+c_{\rm cut})-\frac {c_{\rm cut}}{1+c_{\rm cut}} \right]^{-2},
\eeq

The total $\gamma$-ray intensity at Earth from the annihilation of dark matter particles 
in galactic subhalos can be written after integrating Eq.~\ref{sat:individual} 
over the distribution of Milky Way subhalos as, 
\begin{eqnarray}
\frac{d{\mathcal I}_{\rm sh}(E_{\gamma})}{dE_{\gamma}}&=&\int dVdM \frac{dn_{\rm sh}(M,s,\ell,b)}{dM}\frac{d{\mathcal I}(E_{\gamma},s,M)}{dE_{\gamma}}, 
\label{eq:i_sh}
\end{eqnarray}
where $\int dMdV (dn_{\rm sh}/dM)$ is the total number of subhalos in the Milky Way. 
The form of subhalo mass function, $dn_{\rm sh}/dM$ is chosen to be the 
anti-biased model~\footnote{In another model (`unbiased') for $n_{\rm sh} (r)$, the subhalo
distribution is assumed to follow its parent NFW halo distribution whereas in the anti-biased model
the subhalo distribution is flatter than NFW halo~\citep{Ando:2009fp}.}
following Ref.~\citep{Ando:2009fp} for our calculation.

In order to confront observations, we are interested in the averaged intensity of 
$\gamma$-rays per unit energy emitted due to DM annihilation over the whole galaxy,
\begin{eqnarray}
\label{first}
\frac{dI_{\rm sh}(E_{\gamma})}{dE_{\gamma}} &=& \frac1{\Omega_e} \frac{d{\mathcal I}_{\rm sh}(E_{\gamma})}{dE_{\gamma}} =  \frac1{\Omega_e}  \int_{M_*} \int_{V_*(M)} dVdM
\frac{dn_{\rm sh}(M,s,\ell,b)}{dM}\frac{d{\mathcal I}(E_{\gamma},s,M)}{dE_{\gamma}} \nonumber \\
&=&\int_{M_*} dM \int_{V_*(M)} dV \frac{dn_{\rm sh}(M,s,\ell,b)}{dM} \frac1{4\pi s^2} \frac{\langle \sigma v \rangle}{m_{\rm DM}^2}
\frac{dN_\gamma}{dE_{\gamma}} \frac{M^2}{r_{s,{\rm sh}}(M)^3} {\mathcal G}[c_{\rm cut}(M)]\;
\end{eqnarray}
where $V_*$ is the volume beyond which satellites remain unresolved. 
The considered mass range of the subhalos is 
$10^{-6} M_\odot \leq M_* \leq  10^{10} M_\odot$. Since 
from the luminosity $L$ one gets the knowledge of the subhalo mass $M$
% the subhalo masses ($M$) and luminosities ($L$) are assumed to follow one-to-one relation,
we consider the subhalo mass range in such a way that the bright as well as the faint subhalos
are included in the calculation.  
Also since luminosity is directly related to the flux sensitivity of Fermi ($F_{\rm sens}$) by 
the relation, $L(M) = 4\pi s_*^2(M) F_{\rm sens}$,
they remain unresolvable beyond 
the distance $s_*(M)=\sqrt{L(M)/4\pi F_{\rm sens}}$, 
where the flux sensitivity of Fermi-LAT, 
$F_{\rm sens}=2\times 10^{-10} {\rm cm}^{-2} {\rm s}^{-1}$~\citep{Ando:2009fp} 
and $L(M)$ is the luminosity obtained by integrating Eq.~\ref{sat:lum} over energy.

\begin{figure}[cbt!]
\begin{center}
\includegraphics[angle=-90, scale=0.32]{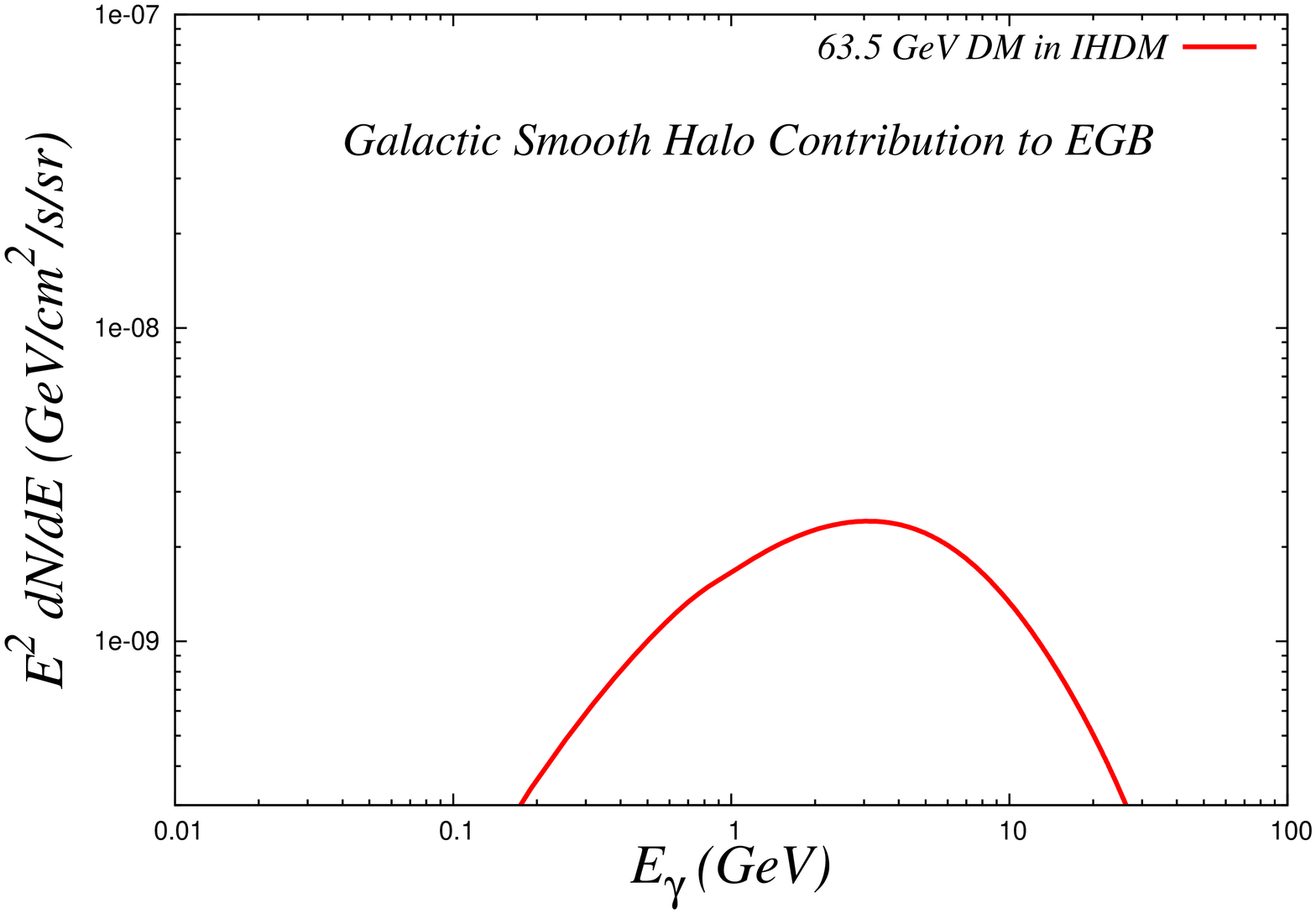}
\includegraphics[angle=-90, scale=0.32]{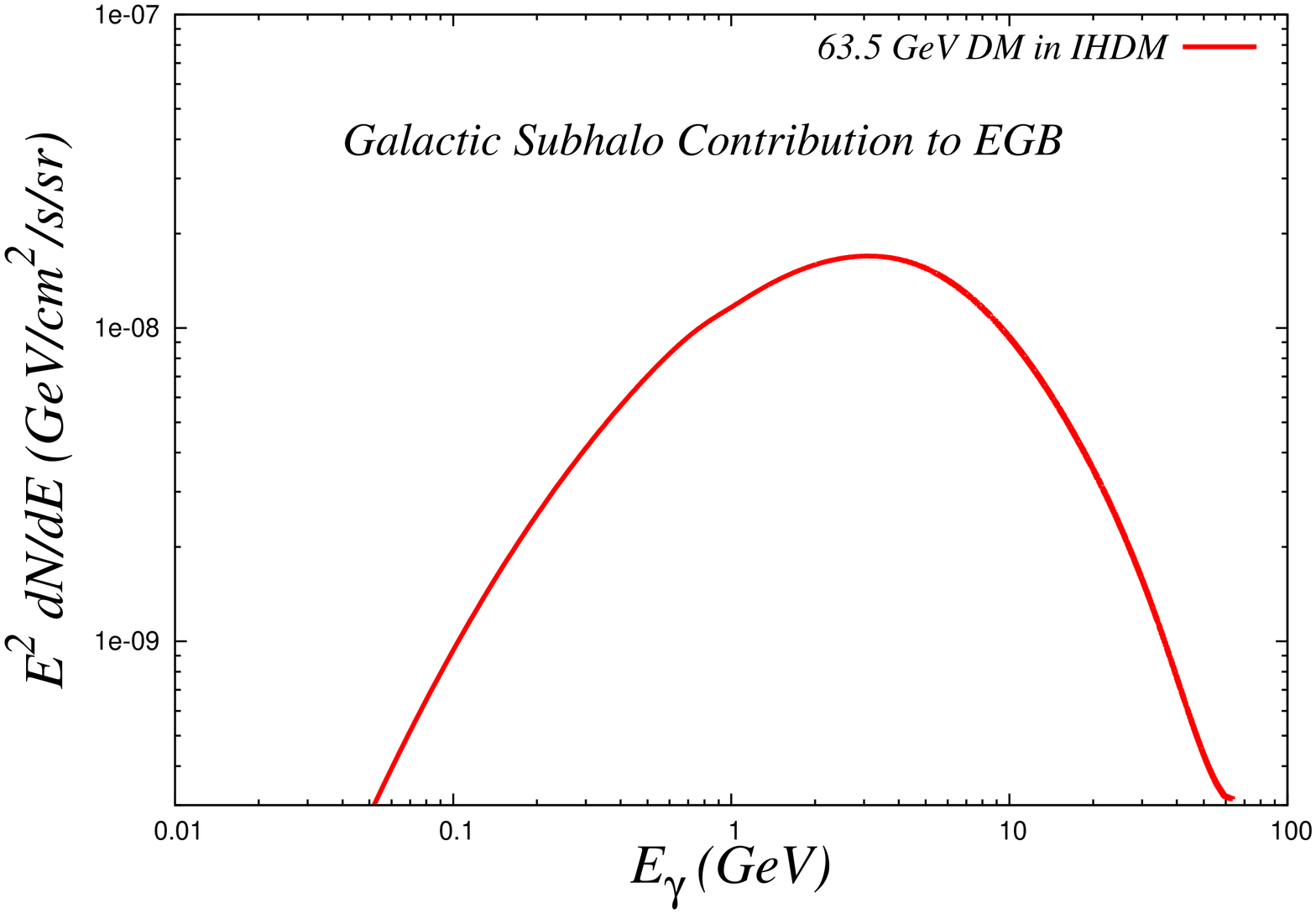}
%%%
   \caption{{\it Galactic smooth halo and subhalo contributions to
   the extragalactic gamma-ray flux for LIP $\sim$ 63.5 GeV dark matter in IHDM. See text for details.}}
   \label{gal_sm_sh}
\end{center}
\end{figure}

In Fig.~\ref{gal_sm_sh}, we show the contributions to the EGB from 
both galactic smooth halos and subhalos. From the left panel of Fig.~\ref{gal_sm_sh}
we see that the contribution to the EGB is much subdominant for the 
annihilation of DM within galactic smooth halo. 
For our chosen galactic substructure model, this contribution is comparable to that from 
extragalactic dark matter annihilations calculated using $c_{vir}$ of Macci\`{o} {\it et al.} model. 
For the other case where the gamma ray flux from 
extragalactic dark matter annihilations have been calculated 
using $c_{vir}$ of power law model, the contributions from the
galactic subhalos are much negligible.

\section{Summary and Conclusion \label{sum_sec}}

We have chosen a simple dark matter (DM) model, namely inert Higgs doublet model (IHDM) where 
scalar sector of Standard Model is extended by adding another SU(2)$_L$ doublet. The newly added
doublet does not generate any VEV after spontaneous symmetry breaking. The `inert' doublet is
considered to be the DM candidate. The stability of DM is ensured by imposing
discrete $\Z_2$ symmetry. 
% an extra 
% complex SU(2)$_L$ doublet is added other than Standard Model Higgs doublet. The stability of
% dark matter particle in this model is ensured by the imposed discrete $\Z_2$ symmetry on this 
% additional doublet. 
The model, in general, provides a broad range of DM mass from GeV
to TeV range. In this study we only consider the lower mass range of DM in this model.
The analysis of experimental data for DM relic density from PLANCK 
experiment and the other direct detection experimental results for the case of this IHDM gives a set
of best fit values for DM mass, annihilation cross section and other model parameters.
We adopt this best fit point (obtained using $\chi^2$ minimisation) for IHDM mass 
from such analyses. Thus the DM mass of $\sim$ 63.5 GeV is our
chosen benchmark point in the present work.
% and choose the best-fit (obtained using $\chi^2$ minimisation) parameter values in this 
% region as our benchmark scenario for further analysis. The best-fit parameter set 
% results a dark matter with mass 63.54 GeV. 
We study the $\gamma$-ray
spectrum obtained from the annihilation of this chosen DM particle
in IHDM framework and interpret various types of continuum $\gamma$-ray fluxes with 
astrophysical origins measured by Fermi-LAT satellite.

In this study we compare our calculated $\gamma$-ray flux with the galactic centre 
$\gamma$-ray excess in the light of this
model. For this we have employed different analysed Fermi-LAT residual $\gamma$-ray flux data 
for different angular regions around the galactic centre. 
% which are performed by several groups using various distinct techniques. 
The calculated low energetic photon 
spectra from the annihilation of the DM particle with benchmark value of mass ($\sim$ 63.5 GeV) in IHDM  
for various chosen regions surrounding the galactic centre are found to be in the same ballpark
as reported by these studies. Although in some previous analyses it was argued that the
photon spectra originated from different
annihilation channels of dark matter particles with low masses can possibly fit the obtained data,
very recent analyses have obtained the resulting best fit masses of dark matter to be 
much more conservative (and also somewhat higher as well).
We have computed the photon spectra for our benchmark scenario in IHDM framework 
and have confronted with the residual photon spectra obtained for all of the above-mentioned
studies. Our theoretical calculations for photon spectra in this model have been performed 
after suitable parametrisation of the dark matter halo parameters, region of interest
surrounding the galactic centre etc. 

We also address the prospects of the continuum $\gamma$-ray signal which may come from
DM-dominated dwarf spheroidal galaxies (dSphs) in case they originate from dark matter annihilation.
We then compare the $\gamma$-ray flux that can be obtained from 
the benchmark IHDM dark matter with mass $\sim$ 63.5 GeV.
For this we choose 18 Milky Way dSphs whose $J$-factor
can be estimated from measurements. The uncertainties in the measurement of $J$-factor
for different dSphs are also incorporated in our calculations. The calculated photon spectra
for IHDM benchmark point are seen to obey the allowed limits 
for observed spectra of continuum $\gamma$-ray.

After addressing the issues regarding indirect DM searches with $\gamma$-ray signals 
from various galactic cases, we finally confront the extragalactic $\gamma$-ray signal 
with that from the annihilation of low mass DM (considered in this work) in IHDM scenario.
We calculate the extragalactic $\gamma$-ray flux for different extragalactic parametrisations
and compare with the observed extragalactic gamma ray background by EGRET and Fermi-LAT.
For this we consider several possible classes of non-DM astrophysical sources which may yield 
$\gamma$-ray signal embedded in the extragalactic background. Although there are too many 
uncertainties involved in modelling of such astrophysical sources and other parameters 
for extragalactic flux calculation, we have shown that the considered low mass DM in IHDM 
can generate photon flux within the observed flux limit.

From the detailed study of various galactic and extragalactic $\gamma$-ray searches for
probing the indirect signatures of DM in light of IHDM, we can conclude 
that the low mass DM
in this model framework is still a viable candidate to be probed in future $\gamma$-ray 
searches. Although we have performed the thorough analysis considering only a single Higgs-portal
model, the analysis is valid for any simple Higgs-portal DM model such as 
singlet scalar DM model, singlet
fermion DM, inert Higgs triplet model etc. 
with DM mass in the same ballpark as in our study. 

% 
% \fi
% 
% \newpage
% %%%%%%%%%%%%%%%%%%%%%%%%%%%%%%%%%%
% \vspace{0.5cm}
% \hspace{0.2cm} {\bf Acknowledgments}
\section*{Acknowledgments}
\vspace{0.5cm}
KPM would like to thank Alexander Pukhov for his help in \texttt{micrOMEGAs} code.
KPM also acknowledge Department of Atomic Energy (DAE, Govt. of India) for financial
assistance.
%%%%%%%%%%%%%%%%%%%%%%%%%%%%%%%%%%
%\clearpage
\bibliography{idm_gamma}

%\clearpage

%%%%%%%%%%%%%%%%%%%%%%%%%%%%%%%%%%%%%%%%%%%%%%%%%%%%%%%%%%%%%%%%%%%%%%%%
%
 \begin{table}[h!]
 \caption{Benchmark Point : IHDM Parameters }
 \centering
% \begin{adjustwidth}{-8mm}
% \hskip -10mm
 \vskip 2 mm
\bgroup
\def\arraystretch{1.5}%
\scalebox{0.9}{
 \begin{tabular}{| c | c | c | c | c | c | c |}
\hline \hline
 Model & $M_{h^0}$ (GeV) & $M_{H^{\pm}}$ (GeV)
& $M_{H^0}$ (GeV) & $M_{A^0}$ (GeV) & $\lambda_L$ & $\lambda_2$  \\
     \cline{2-7}
Parameters & 126.016 & 73.78 & 63.54 & 166.16 & $-3.29\times10^{-3}$ & $5.67\times10^{-4}$ \\   
    \hline \hline
 \end{tabular}}
% \end{adjustwidth}
\egroup
  \label{table1}
\end{table}
%%%%%%%%%%%%%%%%%%%%%%%%%%%%%%%%%%%%%%%%%%%%%%%%%%%%%%%

%%%%%%%%%%%%%%%%%%%%%%%%%%%%%%%%%%%%%%%%%%%%%%%%%%%%%%%
\begin{table}[h!]
 \centering
 \caption{Benchmark Point : Observables}
% \begin{adjustwidth}{-8mm}
% \hskip -10mm
 \vskip 2 mm
\bgroup
\def\arraystretch{1.5}%
 \scalebox{0.84}{
 \begin{tabular}{| c | c | c | c | c | c | c |}
\hline \hline
  DM &   \multicolumn{2}{|c|}{$\Omega h^2$} & \multicolumn{2}{|c|}{$\langle \sigma v \rangle$ (cm$^3$s$^{-1}$)} & \multicolumn{2}{|c|}{$\sigma_{\rm SI}$ (pb)}
     \\
     \cline{2-7}
 Observables & \multicolumn{2}{|c|}{0.1173} & \multicolumn{2}{|c|}{$2.37\times10^{-26}$} & \multicolumn{2}{|c|}{$8.89\times10^{-11}$}
   \\
    \hline \hline
    Annihilation &   \multicolumn{1}{|c|}{$H^0H^0\to b\bar{b}$} & \multicolumn{1}{|c|}{$H^0H^0\to W^+W^-$} & \multicolumn{1}{|c|}{$H^0H^0\to gg$} & \multicolumn{1}{|c|}{$H^0H^0\to l\bar{l}$} & \multicolumn{1}{|c|}{$H^0H^0\to c\bar{c}$} & \multicolumn{1}{|c|}{$H^0H^0\to ZZ$}
     \\
     \cline{2-7}
 Cross-section & \multicolumn{1}{|c|}{69.2\%} & \multicolumn{1}{|c|}{9.61\%} & \multicolumn{1}{|c|}{9.49\%} & \multicolumn{1}{|c|}{7.37\%} & \multicolumn{1}{|c|}{3.29\%} & \multicolumn{1}{|c|}{0.48\%}
   \\
    \hline \hline
 
 \end{tabular} }
% \end{adjustwidth}
\egroup
  \label{table2}
\end{table}

%%#######################################################%%

\clearpage

%%%%%%%%%%%%%%%%%%%%%%%%%%%%%%%%%%%%%%%%%%%%%%%%%%%%%%%%%%%%%%%%%%%%%%%
 \begin{table}[ht!]
\caption{ Limits on DM annihilation cross-section from $\gamma$-ray flux limits for various dwarf spheroidal galaxies for the benchmark DM mass of Table~\ref{table1} }
 \centering
% \begin{adjustwidth}{-8mm}
% \hskip -10mm
%\bgroup
%\def\arraystretch{1.5}%
 \vskip 2 mm
 \scalebox{0.85}{
%\tabletypesize{\tiny}
 \begin{tabular}{|l | c | c | c | c | c | c | c | c |}
\hline
dSphs name  & longitude & latitude & distance & $\overline{log_{10}({\rm J^{NFW}})}$
%~\footnote{$J$-factors are calculated over a solid angle of $\Delta\Omega \sim 2.4 \times 10^{-4}$ sr.} 
& $\overline{log_{10}(\alpha_s ^{\rm NFW})}$ & upper limit on\\
            &  $l$ ($\deg$) &  $b$ ($\deg$) & (kpc) & ($log_{10}[{\rm GeV}^2 {\rm cm}^{-5} {\rm sr}]$) & ($log_{10}[\deg]$) & $\langle\sigma v\rangle$ (cm$^3$s$^{-1}$) \\ \hline\hline
Bootes I                  & 358.1  & 69.6   & 66     & $18.8 \pm 0.22$ & $-0.6 \pm 0.3$ & $2.33\times10^{-24}$ \\ 
\citep{Dall'Ora:2006pt}  & & & & & & \\ \hline
Bootes II   & 353.7  & 68.9   & 42     &              -- &     -- &                  -- \\ \hline
Bootes III  & 35.4   & 75.4   & 47     &              -- &     -- &                   -- \\ \hline
Canes Venatici I & 74.3   & 79.8   & 218    & $17.7 \pm 0.26$ & $-1.3 \pm 0.2$ &  $9.65\times10^{-25}$  \\ 
\citep{Simon:2007dq} & & & & & & \\ \hline
Canes Venatici II & 113.6  & 82.7   & 160    & $17.9 \pm 0.25$ & $-1.1 \pm 0.4$ & $8.14\times10^{-25}$ \\ 
\citep{Simon:2007dq} & & & & & & \\ \hline 
Canis Major    & 240.0  & -8.0   & 7      &              -- &     -- &                   -- \\ 
\citep{Walker:2008ax} &&&&&& \\ \hline 
Carina        & 260.1  & -22.2  & 105    & $18.1 \pm 0.23$ & $-1.0 \pm 0.3$ &  $2.28\times10^{-25}$  \\ 
\citep{Simon:2007dq} &&&&&& \\ \hline 
Coma Berenices     & 241.9  & 83.6   & 44     & $19.0 \pm 0.25$ & $-0.6 \pm 0.5$ & $1.11\times10^{-24}$ \\ 
\citep{Simon:2007dq} &&&&&& \\ \hline 
Draco           & 86.4   & 34.7   & 76     & $18.8 \pm 0.16$ & $-0.6 \pm 0.2$ & $3.87\times10^{-25}$ \\ 
\citep{Munoz:2005be} &&&&&& \\ \hline 
Fornax           & 237.1  & -65.7  & 147    & $18.2 \pm 0.21$ & $-0.8 \pm 0.2$ & $2.53\times10^{-25}$ \\ 
\citep{Walker:2008ax} &&&&&& \\ \hline 
Hercules          & 28.7   & 36.9   & 132    & $18.1 \pm 0.25$ & $-1.1 \pm 0.4$ & $9.97\times10^{-26}$ \\ 
\citep{Simon:2007dq}  &&&&&& \\ \hline 
Leo I            & 226.0  & 49.1   & 254    & $17.7 \pm 0.18$ & $-1.1 \pm 0.3$ & $4.37\times10^{-25}$ \\ 
\citep{Mateo:2007xh} &&&&&& \\ \hline 
Leo II             & 220.2  & 67.2   & 233    & $17.6 \pm 0.18$ & $-1.1 \pm 0.5$ & $3.88\times10^{-25}$ \\
\citep{Koch:2007ye} &&&&&& \\ \hline 
Leo IV           & 265.4  & 56.5   & 154    & $17.9 \pm 0.28$ & $-1.1 \pm 0.4$ & $3.72\times10^{-24}$ \\ 
\citep{Simon:2007dq}  &&&&&& \\ \hline 
Leo V                     & 261.9  & 58.5   & 178    &              -- &     -- &           -- \\ \hline 
Pisces II                 & 79.2   & -47.1  & 182    &              -- &     -- &           -- \\ \hline 
Sagittarius               & 5.6    & -14.2  & 26     &              -- &     -- &           -- \\ \hline 
Sculptor         & 287.5  & -83.2  & 86     & $18.6 \pm 0.18$ & $-0.6 \pm 0.3$ & $3.41\times10^{-24}$\\ 
\citep{Walker:2008ax} &&&&&& \\ \hline 
Segue 1           & 220.5  & 50.4   & 23     & $19.5 \pm 0.29$ & $-0.4 \pm 0.5$ & $1.16\times10^{-24}$ \\ 
\citep{Simon:2010ek}  &&&&&& \\ \hline 
Segue 2                   & 149.4  & -38.1  & 35     &              -- &     -- &           -- \\ \hline 
Sextans         & 243.5  & 42.3   & 86     & $18.4 \pm 0.27$ & $-0.9 \pm 0.2$ & $1.14\times10^{-25}$ \\ 
\citep{Walker:2008ax}   &&&&&& \\ \hline 
Ursa Major I      & 159.4  & 54.4   & 97     & $18.3 \pm 0.24$ & $-1.0 \pm 0.3$ & $1.64\times10^{-25}$ \\
\citep{Simon:2007dq}  &&&&&& \\ \hline 
Ursa Major II       & 152.5  & 37.4   & 32     & $19.3 \pm 0.28$ & $-0.5 \pm 0.4$ & $1.33\times10^{-24}$ \\ 
\citep{Simon:2007dq} &&&&&& \\ \hline 
Ursa Minor       & 105.0  & 44.8   & 76     & $18.8 \pm 0.19$ & $-0.5 \pm 0.2$ & $6.54\times10^{-24}$ \\ 
\citep{Munoz:2005be}  &&&&&& \\ \hline 
Willman 1       & 158.6  & 56.8   & 38     & $19.1 \pm 0.31$ & $-0.6 \pm 0.5$ & $4.03\times10^{-24}$ \\ 
\citep{Willman:2010gy} &&&&&& \\ \hline 
\end{tabular}}
%\egroup
\label{table3}
\end{table}
%%%%%%%%%%%%%%%%%%%%%%%%%%%%%%%%%%%%%%%%%%%%%%%%%%%%%%%%%%%%%%%%%%%%%%

\clearpage

%%%%%%%%%%%%%%%%%%%%%%%%%%%%%%%%%%%%%%%%%%%%%%%%%%%%%%%%%%%%%%%%%%%%%%%%
%
 \begin{table}[h!]
 \caption{Overview of the minimal non-DM contributions to the total extragalactic $\gamma$-ray background~\citep{Tavakoli:2013zva}}
 \centering
% \begin{adjustwidth}{-8mm}
% \hskip -10mm
 \vskip 2 mm
\bgroup
\def\arraystretch{1.5}%
 \begin{tabular}{| c | c | c |}
\hline \hline
 Non-DM objects & Photon Energy Spectra (${dN \over dE}$ in $\textrm{GeV}^{-1}\textrm{cm}^{-2}\textrm{s}^{-1}\textrm{sr}^{-1}$) \\
\hline \hline
BL Lacs & $3.9\times 10^{-8} E_{\gamma}^{-2.23}$  \\ 
\hline   
FSRQ & $3.1\times 10^{-8} E_{\gamma}^{-2.45}$  \\
\hline
MSP & $1.8\times 10^{-7} 
        E_{\gamma}^{-1.5} \exp\left(-{E_{\gamma}\over 1.9}\right)$ \\
\hline 
SFG & $1.3\times 10^{-7} 
        E_{\gamma}^{-2.75}$ \\
\hline
FR~I \& FR~II & $5.7\times 10^{-8} 
        E_{\gamma}^{-2.39} \exp\left(-{E_{\gamma}\over 50.0}\right)$ \\
\hline
UHECR & $4.8\times 10^{-9} 
        E_{\gamma}^{-1.8} \exp\left[-\left({E_{\gamma}\over 100.0}\right)^{0.35}\right]$ \\
\hline 
GRB & $8.9\times 10^{-9} 
        E_{\gamma}^{-2.1}$ \\
\hline
SBG & $0.3\times 10^{-7} 
        E_{\gamma}^{-2.4}$ \\
\hline 
UHEp ICM & $3.1\times 10^{-9} 
        E_{\gamma}^{-2.75}$  \\
\hline
IGS & $0.87 \times 10^{-10} \times\left\{\begin{array}{ll} 
	 & \left(\frac{E_{\gamma}}{10}\right)^{-2.04} \; \textrm{for} 
	 \; E_{\gamma} < 10\ \textrm{GeV} \\
	 & \left(\frac{E_{\gamma}}{10}\right)^{-2.13} \; \textrm{for} \; E_{\gamma} > 10\ \textrm{GeV}
	 \end{array}\right\}$ \\
    \hline \hline
 \end{tabular}
\egroup
% \end{adjustwidth}
  \label{table4}
\end{table}
%%%%%%%%%%%%%%%%%%%%%%%%%%%%%%%%%%%%%%%%%%%%%%%%%%%%%%%%%%%%%%%%%%%%%%%%%
\clearpage

%%%%%%%%%%%%%%%%%%%
%%%%%%%%%%%%%%%%%%%

%%%%%%%%%%%%%%%%%%%
%%%%%%%%%%%%%%%%%%%

\clearpage

%%%%%%%%%%%%%%%%%%%
%%%%%%%%%%%%%%%%%%%

%\clearpage

%%%%%%%%%%%%%%%%%%%
%%%%%%%%%%%%%%%%%%%

\clearpage

%%%%%%%%%%%%%%%%%%%
%\begin{figure}[ht!]
%%%%%%%%%%%%%%%%%%%

\clearpage

%%%%%%%%%%%%%%%%%%%
%%%%%%%%%%%%%%%%%%%

%%%%%%%%%%%%%%%%%%%
%%%%%%%%%%%%%%%%%%%

\clearpage

%%%%%%%%%%%%%%%%%%%
%%%%%%%%%%%%%%%%%%%

%%%%%%%%%%%%%%%%%%%
\begin{figure}[cbt!]
\begin{center}
\includegraphics[angle=-90, scale=0.32]{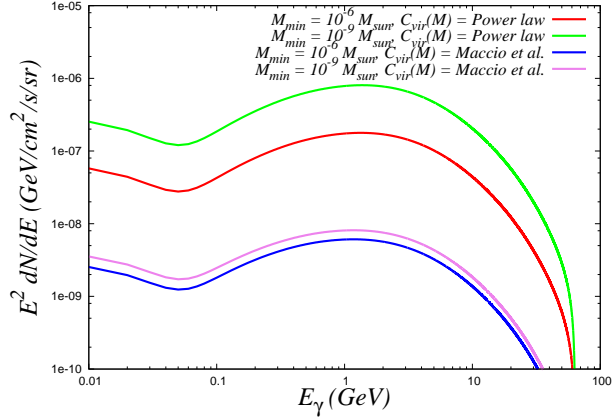}
%%%
   \caption{{\it Comparison of the computed extragalactic gamma-ray fluxes from the annihilation of 
   $\sim$ 63.5 GeV dark matter (benchmark point) in IHDM framework for different 
   extragalactic parametrisations. We consider two models of the concentration parameter $c_{vir}$,
   namely a)power law model and b)Macci\`{o} {\it et al.} model. Also the minimum extragalactic
   subhalo mass $M_{\rm min}$ are chosen to be 
   $10^{-6} M_{\odot}\,\,{\rm and}\,\, 10^{-9}\, M_{\odot}$. Calculation with 
   power law model yields enhanced $\gamma$-ray flux compared to that with Macci\`{o} {\it et al.} 
   model. For low $M_{\rm min}$, $\gamma$-ray flux increases and this enhancement is
   smaller for Macci\`{o} {\it et al.} model compared to that for power law model. See text 
   for details.}}
   \label{exgal}
\end{center}
\end{figure}
%%%%%%%%%%%%%%%%%%%
\clearpage

%%%%%%%%%%%%%%%%%%%
%%%%%%%%%%%%%%%%%%%
\clearpage

%%%%%%%%%%%%%%%%%%%
%%%%%%%%%%%%%%%%%%%

%%%%%%%%%%%%%%%%%%%
%%%%%%%%%%%%%%%%%%%

\end{document}